\documentclass[lettersize,journal]{IEEEtran}
\IEEEoverridecommandlockouts
\usepackage{cite}
\usepackage{hyperref}
\usepackage{amsmath,amssymb,amsfonts}
\usepackage{algpseudocode}
\usepackage{algorithm}
\usepackage{algorithmicx}
\usepackage{graphicx}
\usepackage{textcomp}
\usepackage{xcolor}
\usepackage{enumitem}
\usepackage{subcaption}
\usepackage{flushend}

\begin{document}

\title{
Trajectory-Aware Multi-RIS Activation and Configuration: A Riemannian Diffusion Method
%
}

\author{ 
Kaining~Wang,~\IEEEmembership{Student Member,~IEEE,} 
Bo~Yang,~\IEEEmembership{Senior Member,~IEEE,} Yusheng Lei, Zhibo Li,\\ Zhiwen Yu,~\IEEEmembership{Senior Member,~IEEE,} Xuelin Cao,~\IEEEmembership{Senior Member,~IEEE,} Bin Guo,~\IEEEmembership{Senior Member,~IEEE,}\\ George C. Alexandropoulos, \IEEEmembership{Senior Member,~IEEE,} Dusit Niyato,~\IEEEmembership{Fellow,~IEEE},\\ M\'erouane Debbah,~\IEEEmembership{Fellow,~IEEE}, and Zhu Han,~\IEEEmembership{Fellow,~IEEE}
 \thanks{K. Wang, B. Yang, Y. Lei, and B. Guo are with the School of Computer Science, Northwestern Polytechnical University, Xi'an, Shaanxi, 710129, China (email: wangkaining@mail.nwpu.edu.cn, yang$\_$bo, guob@nwpu.edu.cn, ). 

 Z. Li is with the School of Automation, Northwestern Polytechnical University, Xi'an, Shaanxi, 710129, China (email: lizhibo@mail.nwpu.edu.cn). 

Z. Yu is with the School of Computer Science, Northwestern Polytechnical University, Xi'an, Shaanxi, 710129, China, and Harbin Engineering University, Harbin, Heilongjiang, 150001, China (email: zhiwenyu@nwpu.edu.cn).

 X. Cao is with the School of Cyber Engineering, Xidian University, Xi'an, Shaanxi, 710071, China (email: caoxuelin@xidian.edu.cn). 


 G. C. Alexandropoulos is with the Department of Informatics and Telecommunications, National and Kapodistrian University of Athens, 16122 Athens, Greece (email: alexandg@di.uoa.gr). 

D. Niyato is with the College of Computing and Data Science, Nanyang Technological University, Singapore (dniyato@ntu.edu.sg).


M. Debbah is with  KU 6G Research Center, Department of Computer and Information Engineering, Khalifa University, Abu Dhabi 127788, UAE (email: merouane.debbah@ku.ac.ae).

Z. Han is with the Department of Electrical and Computer Engineering at the University of Houston, Houston, TX 77004 USA, and also with the Department of Computer Science and Engineering, Kyung Hee University, Seoul, South Korea, 446-701(email: hanzhu22@gmail.com).

}
}

\maketitle

\begin{abstract}
Reconfigurable intelligent surfaces (RISs) offer a low-cost, energy-efficient means for enhancing wireless coverage. Yet, their inherently programmable reflections may unintentionally amplify interference, particularly in large-scale, multi-RIS-enabled mobile communication scenarios where dense user mobility and frequent line-of-sight overlaps can severely degrade the signal-to-interference-plus-noise ratio (SINR). To address this challenge, this paper presents a novel generative multi-RIS control framework that jointly optimizes the ON/OFF activation patterns of multiple RISs in the smart wireless environment and the phase configurations of the activated RISs based on predictions of multi-user trajectories and interference patterns. We specially design a long short-term memory (LSTM) artificial neural network, enriched with speed and heading features, to forecast multi-user trajectories, thereby enabling reconstruction of future channel state information. To overcome the highly nonconvex nature of the multi-RIS control problem, we develop a Riemannian diffusion model on the torus to generate geometry-consistent phase-configuration, where the reverse diffusion process is dynamically guided by reinforcement learning. We then rigorously derive the optimal ON/OFF states of the metasurfaces by comparing predicted achievable rates under RIS activation and deactivation conditions. Extensive simulations demonstrate that the proposed framework achieves up to 30\% SINR improvement over learning-based control and up to 44\% gain compared with the RIS always-on scheme, while consistently outperforming state-of-the-art baselines across different transmit powers, RIS configurations, and interference densities.\footnote{The source code will be fully released upon acceptance at \url{https://github.com/wangxn2/TPGC}.}
\end{abstract}


\section{Introduction}
\begin{figure}[t]
    \centering
    \includegraphics[width=0.9\columnwidth]{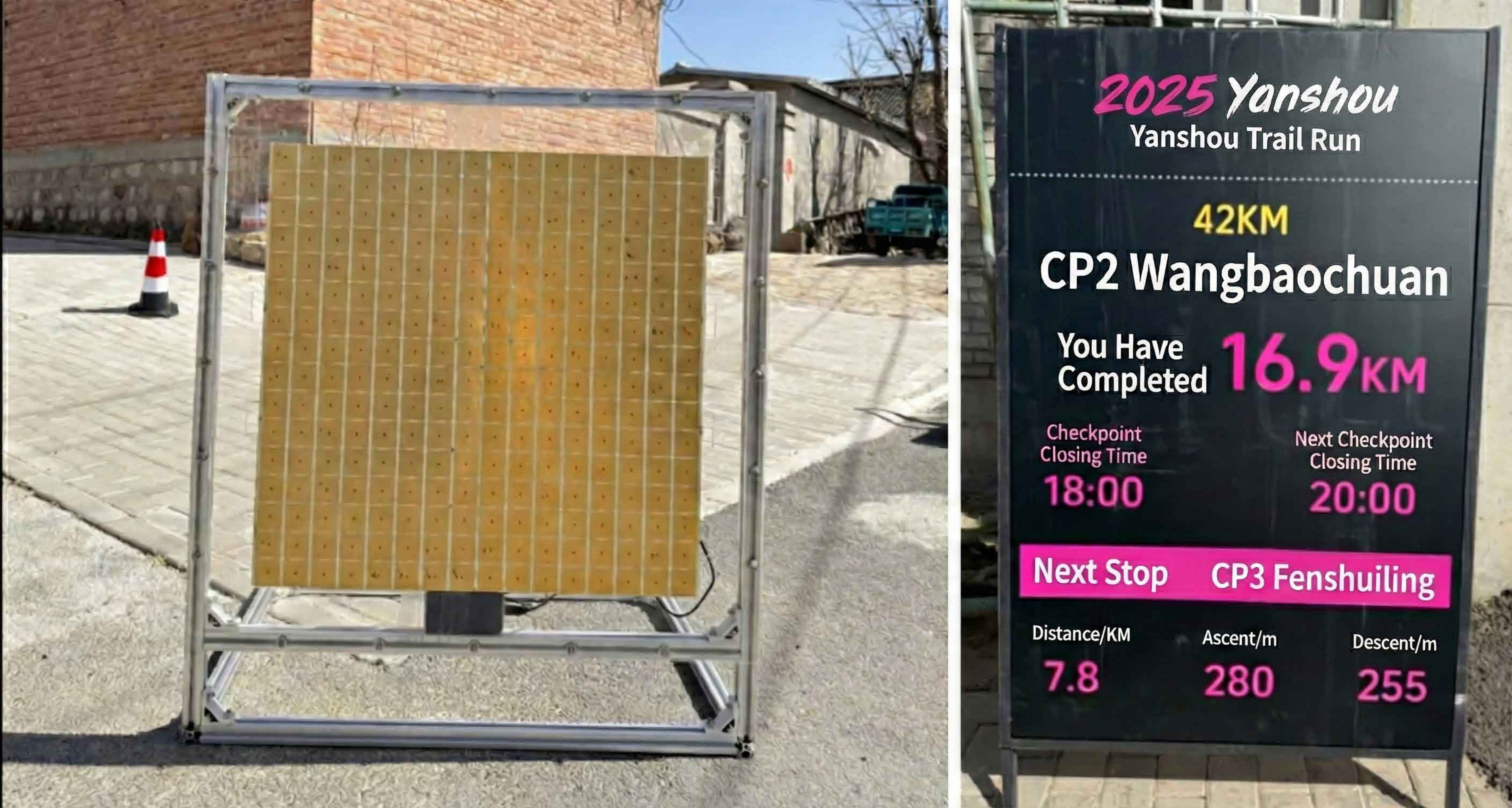}
    \captionsetup{font={small}}
    \caption{On-site deployment of a 3.5 GHz RIS during the Beijing Changping Yanshou Trail Challenge 2025~\cite{beijing2025news}, providing enhanced wireless coverage in marathon scenarios.}
    \label{fig:USE}
\end{figure}

\IEEEPARstart{W}{ith} the rapid growth of large-scale smart urban environments, such as marathons, crowded streets, shopping malls, and large outdoor gatherings, have become increasingly common in next-generation wireless networks~\cite{R19}. In these scenarios, numerous mobile users equipped with smart devices generate highly dynamic traffic patterns, frequent spectrum overlaps, and rapidly varying interference conditions. Ensuring reliable uplink communication in such mobility-intensive and interference-prone environments remains a fundamental challenge.
Ongoing research on sixth-generation (6G) wireless networks introduces reconfigurable intelligent surfaces (RISs) to address these gaps~\cite{10596064,risRoadmap}, and the metasurface is composed of numerous low-cost passive reflecting elements. These passive elements intelligently manipulate the phase and amplitude of incident electromagnetic waves to enhance coverage. This process enables virtual line-of-sight (LoS) paths and reduces signal blockages~\cite{RIS_challenges}. 
A notable example is 
the 2025 Beijing Changping Yanshou Trail Challenge, as shown in Fig.~\ref{fig:USE}. Engineers deployed a 3.5 GHz RIS along the race route, and the devices successfully eliminated coverage gaps. This test demonstrated the effectiveness of RISs in marathon scenarios \cite{beijing2025news}, where they served as auxiliary programmable reflectors to enhance uplink performance, thereby supporting athlete safety.

Despite these advantages, the passive and blind nature of RIS reflection faces critical challenges. Since RIS unit elements reflect incident signals indiscriminately, they not only enhance desired uplink signals, but may also inadvertently reflect interfering signals \cite{tang2021wireless,10729719}. This problem intensifies when multiple users are in proximity within similar angular regions or LoS alignment. The resulting RIS reflections can cause spectral pollution, amplify interference, and severely degrade the signal-to-interference-plus-noise ratio (SINR). 

To mitigate interference amplification and improve the robustness of multi-RIS-assisted communication systems, this paper investigates the joint optimization of the multi-RIS ON/OFF activation pattern and the phase configurations of the activated metasurfaces. The system controls how an RIS reflects signals and decides whether it reflects them at all. This strategy eliminates harmful reflections but preserves beneficial LoS links. Specifically, we propose a trajectory-prediction-based generative ON/OFF control framework (TPGC), which generates high-quality RIS phase configurations while simultaneously determining optimal switching states for interference mitigation. 
Our contributions are outlined as follows:
\begin{itemize}
\item Users in dense mobile environments often follow structured or semi-predictable trajectories. These users frequently operate in the near-field region of multiple RIS panels. Unfortunately, blind RIS reflections can severely amplify co-channel interference~\cite{RIS_challenges}. To proactively suppress such interference, we propose a trajectory-predicted RIS control strategy that uses a long short-term memory (LSTM) to predict mobility. This predictor forecasts the future locations of both target and interfering users. By reconstructing future cascaded channel state information (CSI) from predicted trajectories, the base station (BS) can anticipate harmful geometric overlaps across the RIS Fresnel regions. Consequently, the system dynamically determines the optimal RIS ON/OFF states and phase configurations, thereby facilitating interference mitigation and stable uplink transmission.

\item To address the mixed discrete-continuous and highly non-convex nature of the dynamic RIS response control problem, we formulate the RIS configuration as a conditional generative optimization problem. We then develop a Riemannian diffusion model (RDM) to solve this problem. This model operates directly on the torus manifold of RIS phases. Unlike existing Euclidean diffusion, it injects and removes noise within the phase manifold, thereby preserving the intrinsic periodicity and geometric consistency of the phase shifts. Furthermore, a deep reinforcement learning (DRL)-based critic guides this diffusion process, enabling efficient exploration of high-quality phase solutions without requiring labels.

\item We perform extensive simulations using real-world GeoLife trajectory data to evaluate the proposed framework. The results demonstrate that the proposed method consistently outperforms representative RIS control schemes. It also surpasses state-of-the-art DRL and diffusion-based baselines in terms of SINR. Furthermore, the framework exhibits superior robustness to user mobility and better generalization across interference densities. We conducted additional ablation studies across different system configurations to further validate the stability and practicality of the proposed approach.
\end{itemize}

In this paper, Section~\ref{s2} provides an overview of related works. Section~\ref{s3} presents the system model and problem formulation. Section~\ref{s4} presents the overall TPGC framework. Section~\ref{s5} presents the LSTM-based trajectory prediction model along with the enhanced feature design. Section~\ref{s6} describes the proposed RDM with DRL-based phase optimization for joint RIS phase and switching control. Section~\ref{s7} presents simulation results. In conclusion, Section~\ref{s8} concludes the paper.

\textit{Notations:} Plain symbols denote scalars (e.g., $a$, $b$, $d_u$), bold lowercase symbols denote vectors (e.g., $\boldsymbol{\Phi}$, $\mathbf{x}$), bold uppercase symbols denote matrices or vector-valued functions (e.g., $\mathbf{V}$, $\mathbf{H}$), and calligraphic symbols denote sets (e.g., $\mathcal{U}$, $\mathcal{R}$). The operator $\|\cdot\|$ (or $\|\cdot\|_2$) denotes the Euclidean norm, while $(\cdot)^{+}$ represents $\max\{0,\cdot\}$. $\mathbb{E}[\cdot]$ denotes the expectation operator, and $\mathcal{N}(\mathbf{0},\mathbf{I})$ denotes the standard multivariate Gaussian distribution. The functions $\exp(\cdot)$, $\log_2(\cdot)$, and $\mathrm{atan2}(\cdot,\cdot)$ denote the exponential, base-2 logarithm, and four-quadrant inverse tangent functions, respectively. $(S^1)^N$ denotes the $N$-dimensional torus manifold of RIS phase shifts, $\Pi_{\mathcal{T}}(\cdot)$ denotes the projection onto the torus manifold.

\section{Related Work}\label{s2}
\subsection{RIS Activation Control}
RIS control has attracted significant research interest, particularly in designing ON/OFF state and passive phase configurations that maximize link quality. Early studies focused primarily on optimizing phase shifts under idealized, interference-free assumptions that do not accurately reflect practical multi-user environments. For example, Xie and Li \cite{tcomm} investigated the fundamental problem of whether an RIS should reflect signals or remain inactive based on a binary ON/OFF decision rule, but their analysis was confined to static channels and did not account for external interference. Zheng et al. \cite{10473082} similarly developed a Huygens–Fresnel–based phase configuration strategy for a 1-bit RIS. This approach demonstrated clear near-field geometric benefits. However, the method relies on precise geometric information, assumes stable, quasi-static users, and fails to account for dynamic mobility.

Recognizing the importance of interference, Yang et al. \cite{tvt} introduced intelligent spectrum learning (ISL). This system employs a convolutional neural network (CNN) in an RIS controller to distinguish between desired and interfering signals and adaptively configure their ON/OFF states. While ISL is among the first attempts to integrate interference awareness into RIS activation switching, its system model assumes static user positions and overlooks mobility, trajectory dynamics, and associated prediction uncertainty. As a result, existing switching and optimization methods, despite their contributions, do not fully address the coupled challenges of interference, mobility, and near-field signal behavior that arise in real-world multi-RIS deployment scenarios.

\subsection{DRL via Diffusion Models for Network Optimization}
DRL has become a powerful paradigm for addressing complex decision-making problems in dynamic wireless environments. In particular, DRL has been successfully applied to jointly optimize RIS phase configurations and active beamforming in dynamic propagation environments. For example, Huang \emph{et al.} \cite{drlRIS2} proposed a DRL-based hybrid beamforming framework for multi-hop RIS-empowered terahertz communications, where the actor–critic architecture is leveraged to jointly optimize the digital beamforming at the BS and the analog phase shifts of multiple cascaded RISs under severe path loss and multi-hop channel coupling. In a broader smart radio environment setting, the authors in \cite{drlRIS1} demonstrated that DRL-based control policies can effectively adapt RIS configurations to time-varying channels and network dynamics without relying on explicit analytical models. However, most DRL policies generate deterministic or unimodal Gaussian actions, thereby limiting their capacity to model the inherently multimodal and highly nonconvex action spaces in RIS control problems.

Meanwhile, diffusion models \cite{yang2025frontiers} have recently emerged as prominent generative tools for addressing non-convex optimization tasks in wireless systems. Built upon denoising diffusion probabilistic models (DDPMs)~\cite{ho2020denoising}, diffusion frameworks learn to sample from complex target distributions by gradually injecting noise and reversing the diffusion process via a learned denoising network. In wireless networks, early diffusion-based methods such as GDMOPT~\cite{GDMOPT} demonstrated the feasibility of using diffusion models to learn shared structural priors for combinatorial optimization. At the same time, DFSS~\cite{dfss} applied a hybrid discrete-continuous diffusion mechanism to secure integrated sensing and communications (ISAC) topologies. To address complex optimization problems while reducing reliance on optimal solution data, Liang et al.~\cite{liang2025gdsg} introduced the graph diffusion-based solution generator (GDSG), which leverages graph diffusion techniques to generate candidate solutions for network optimization, with a focus on multi-access edge computing (MEC) networks.

\begin{figure*}[t]
    \centering
    \captionsetup{font={small}}
    \includegraphics[width=0.8\textwidth]{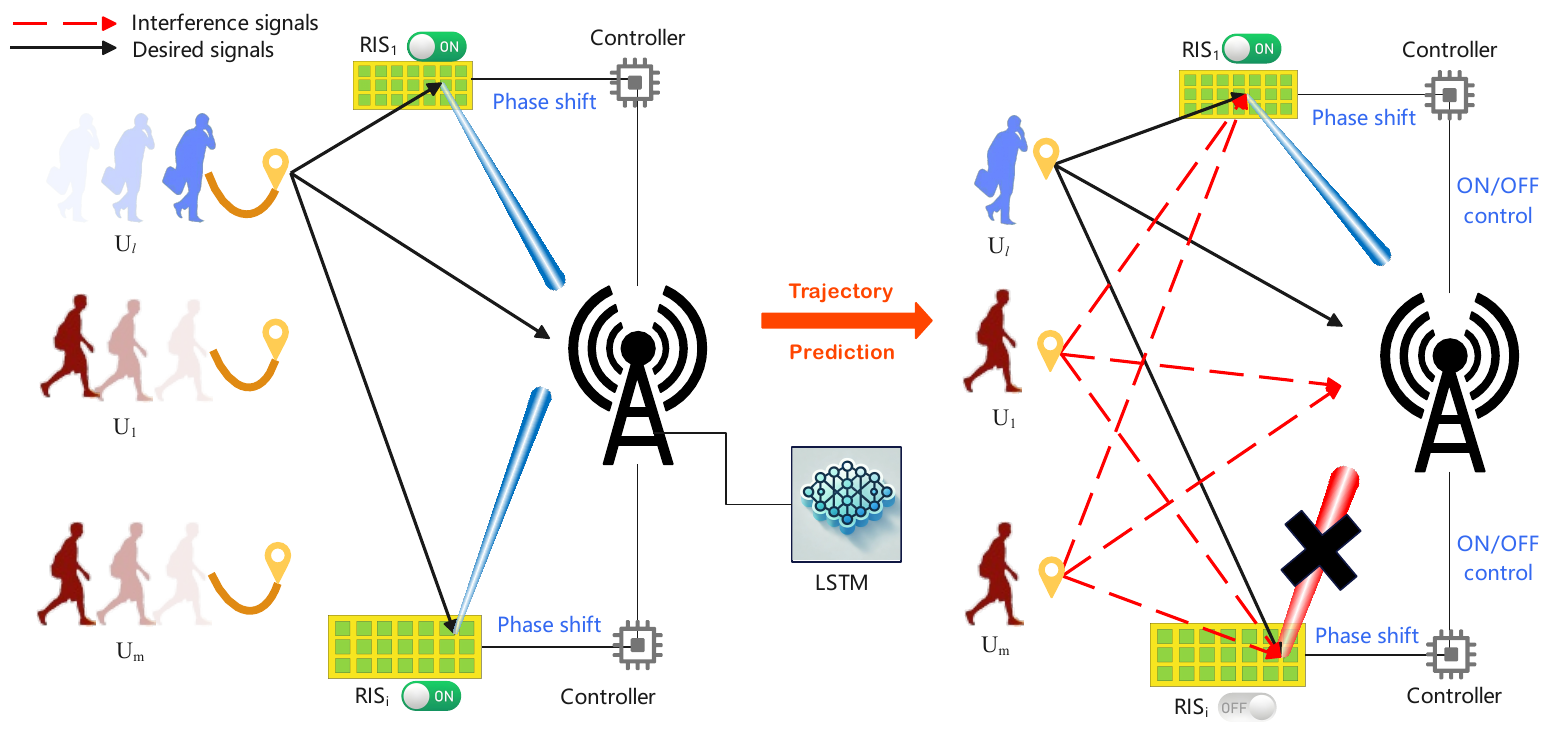}
    \caption{The RIS reflects both desired and interfering signals, resulting in potential performance degradation. With the TPGC framework, BS predicts user trajectories and dynamically manages RIS ON/OFF states, enhancing signal quality and minimizing interference in rapidly changing dense mobile environments.}
    \label{fig:modelDesign}
\end{figure*}

Recent efforts have increasingly integrated diffusion models with DRL to improve action exploration and robustness. For instance, diffusion-augmented reinforcement learning frameworks, such as D2SAC~\cite{du2024diffusion}, G-DDPG~\cite{du2024reinforcement}, and DSAC-QMIX~\cite{ning2025diffusion}, have shown that diffusion-driven policies can generate more diverse and higher-quality actions than traditional DRL policies, thereby improving performance for multi-agent resource management and distributed wireless control tasks. 
In the context of RIS-assisted systems, diffusion models have been explored across multiple layers of system design. Wang et al.~\cite{wang2025optimization} formulated large-scale three-dimensional RIS deployment as a conditional generation problem, in which a DDPM-based generator learns near-optimal deployment patterns under geometric constraints. Zhang et al.~\cite{zhang2025decision} combined a diffusion-based CSI generator with a decision transformer to enable robust, data-driven RIS beamforming from offline trajectories. Ni et al.~\cite{ni2025conditional} further proposed a conditional diffusion model to reconstruct cascaded channels in dual-RIS systems, leveraging spatial correlation to reduce pilot overhead. These works highlight the significant potential of diffusion models for RIS-aided communications, both as generative optimizers and as mechanisms for CSI completion.

However, the existing studies rely on Euclidean diffusion processes with Gaussian perturbations, which implicitly treat RIS phases as unconstrained real-valued variables. This overlooks the intrinsic periodicity and manifold structure of phase shifts, leading to suboptimal or physically inconsistent solutions. Motivated by these limitations, this paper develops an RDM that performs noise injection and denoising directly on the torus manifold, enabling geometry-consistent phase generation and seamless integration with DRL guidance for joint RIS phase configuration and activation control.

\section{System Model and Problem Formulation} \label{s3}
\subsection{RISs-aided Network Scenario}
We consider an RIS-assisted uplink communication system comprising one BS, $R$ RISs, and $U$ mobile users, where $R \ge 1$ and $U \ge R$ typically hold. Let $\mathcal{R}$ and $\mathcal{U}$ denote the index sets of RISs and users, respectively, with each RIS indexed by $\mathrm{RIS}_i$, $i \in \mathcal{R}$, and each user indexed by $u \in \mathcal{U}$. 
As illustrated in Fig.~\ref{fig:modelDesign}, we partition the user set $\mathcal{U}$ into two subsets: a set of \emph{desired users} $\mathcal{U}_l$ and a set of \emph{interfering users} $\mathcal{U}_m$.
Specifically, $\mathcal{U}_l$ denotes the users whose uplink transmissions are of primary interest and are explicitly supported by the RIS-assisted design, whereas $\mathcal{U}_m$ comprises all remaining users that share the uplink spectrum.
Accordingly, the interfering users are defined as the complement of the desired users, i.e., 
 $\mathcal{U}_l \bigcup \mathcal{U}_m = \mathcal{U}$ and $\mathcal{U}_l \bigcap \mathcal{U}_m = \varnothing$.
This formulation captures general co-channel interference in dense mobile environments, where interference originates from all non-target users.

\begin{figure}[t]
    \centering
    \captionsetup{font={small}}
    \includegraphics[width=0.95\linewidth]{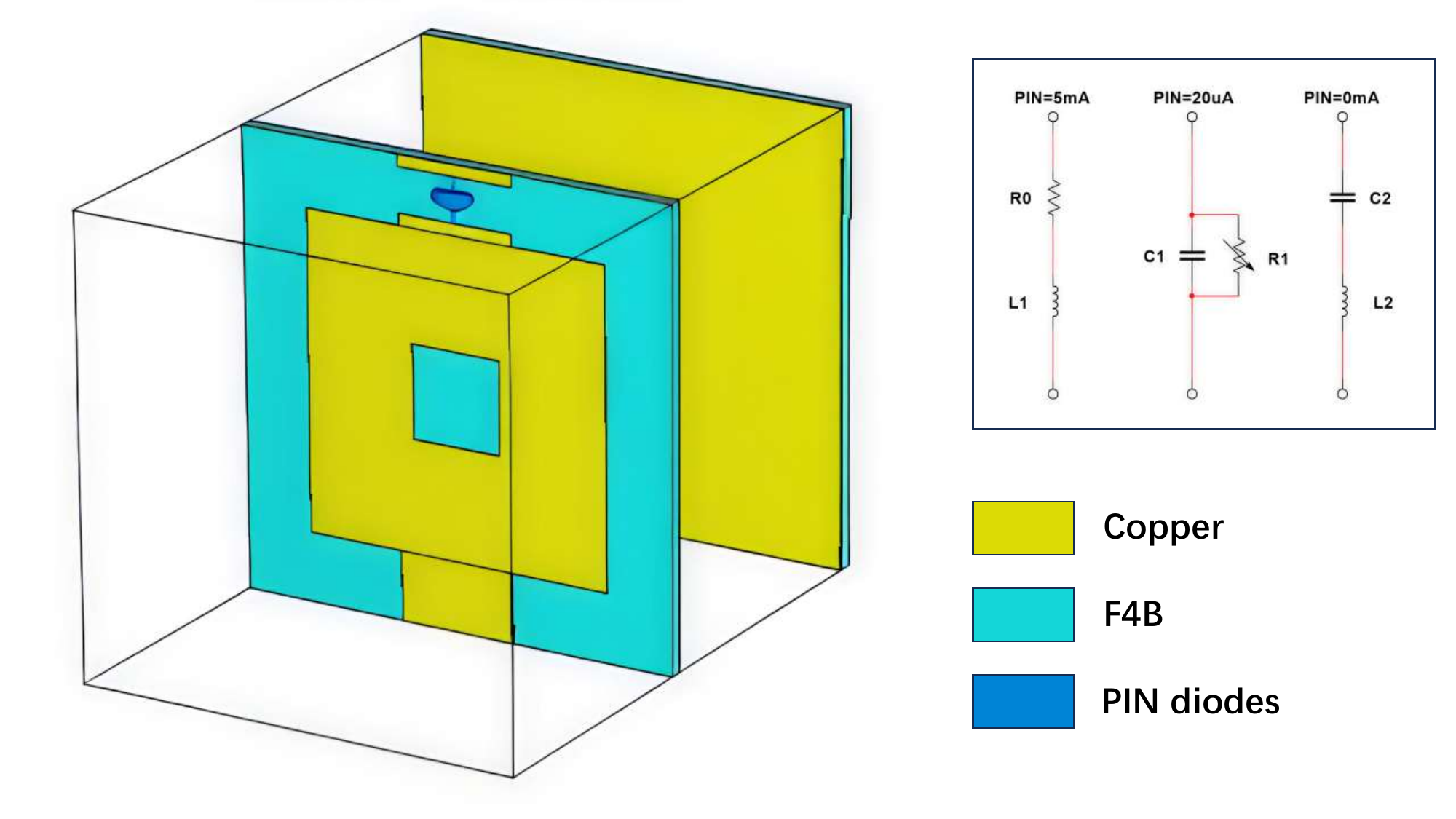}
    \caption{The proposed RIS unit cell for ON/OFF states. The left figure is the RIS structure with copper on an F4B substrate and embedded PIN diodes. The right figure shows the corresponding equivalent diode circuits under different PIN bias currents.}
    \label{fig:structure}
\end{figure}

Without loss of generality, we consider that each RIS primarily assists at most one desired user and operates in two states: ON (activated) or OFF (deactivated). Fig.~\ref{fig:structure} shows one unit cell of the proposed RIS, which is loaded with PIN diodes controlled by bias currents \cite{reconfigurable2021wanga}. The corresponding effective circuits under different diode regimes are illustrated in the upper-right corner of Fig.~\ref{fig:structure}. In the ON state, the PIN diodes are biased between $0\mathrm{mA}$ and $5\mathrm{mA}$, under which the metasurface behaves as a reflective element. In the OFF state, the PIN diodes are weakly biased at $20\mu\mathrm{A}$ so that the unit cell operates in an absorbing mode \cite{absorptiveRIS1,absorptiveRIS2}, i.e., the incident wave is largely dissipated rather than reradiated. Therefore, the OFF state can be modeled as the absence of an RIS-assisted reflection link.
We introduce a binary control variable $v_i\in\{0,1\}$ to represent the operational state of $\mathrm{RIS}_i$: $v_i=1$ indicates the reflective (ON) state, while $v_i=0$ indicates the absorptive (OFF) state.

In the considered dense multi-user wireless environments, a single RIS can simultaneously affect the channels of multiple users by controlling the propagation environment and passive reflections \cite{liu2020reconfigurable}.
Accordingly, the presence of many users within the electromagnetic influence region of the same RIS does not preclude effective RIS-assisted uplink control.
When the number of users exceeds the number of RISs, the BS can employ standard multiple-access and user selection mechanisms (e.g., TDMA, OFDMA~\cite{TDMA}) to allocate shared time--frequency resources among users, while jointly optimizing RIS configurations to enhance desired links and suppress interference \cite{MISO}.
The proposed RIS ON/OFF control further mitigates interference by disabling particular RIS panels when they would exacerbate it.

We assume that each RIS is equipped with $N$ reflecting elements, each of which enables continuous phase tuning. All the RISs are centrally coordinated by the BS, following the RIS-compatible control-plane architecture~\cite{RIScontrol}. The BS acts as the network decision-maker and explicitly configures the RIS behavior via control signaling. In this framework, the RIS controller receives dedicated instructions from the BS to adjust its ON/OFF states, to adapt to the time-varying wireless environment. In particular, the proposed method employs a trajectory-prediction mechanism~\cite{ON-OFF} 
to estimate users' future locations and the spatial distribution of interference. Based on these predictions, the BS can pre-configure RIS ON/OFF states to improve the SINR.


\subsection{Channel Model}
In the scenarios considered, the communication environment is typically LoS-dominant and has limited static blockages along the road. By densely deploying RIS panels along the road (e.g., on lampposts, temporary infrastructure, and roadside facilities), virtual LoS links can be effectively established among users, RISs, and the BS. Therefore, we adopt a deterministic LoS geometric channel model, where the channel coefficient is characterized by large-scale path loss and distance-dependent phase rotation \cite{yuanIndoorRISassistedWireless2024a}.
Specifically, the direct channel between user $u$ and the BS is modeled as
\begin{equation}\label{eq:h_u}
h_{u}=d_u^{-\alpha/2}
\exp\left(-j\frac{2\pi }{\lambda}d_u\right),
\end{equation}
where $d_u$ is the distance between user $U_u$ and the BS, $\alpha$ is the path-loss exponent, and $\lambda$ denotes the carrier wavelength.

Let $d_i$ denote the distance between $\mathrm{RIS}_i$ and the BS, and let $d_{ui}$ denote the distance between user $U_u$ and $\mathrm{RIS}_i$. The reflected link from user $U_u$ to $\mathrm{RIS}_i$ is modeled as
\begin{equation}\label{eq:h_ui}
h_{ui} =
\frac{1}{d_{ui}^{\alpha/2}}
\exp\left(-j\frac{2\pi}{\lambda} d_{ui}\right),
\end{equation}
and the reflected link from $\mathrm{RIS}_i$ to the BS is modeled as
\begin{equation}\label{eq:h_i}
h_{i} =
\frac{1}{d_{i}^{\alpha/2}}
\exp\left(-j\frac{2\pi}{\lambda} d_{i}\right).
\end{equation}

The signal received by the BS from the desired user $U_l$ is given by
\begin{equation}
y_l=\left(h_l+\sum_{i=1}^{R} v_i \boldsymbol{\Phi}_i h_{i} h_{li}
\right)\sqrt{P}\,+ I + \sigma,
\end{equation}
where $\sigma$ is the additive white Gaussian noise (AWGN), and the reflection matrix of the $i$th RIS is defined as
\begin{equation}
\boldsymbol{\Phi}_i = \mathrm{diag}\!\left(e^{j\theta^i_1},e^{j\theta^i_n},\ldots,e^{j\theta^i_N}\right),
\end{equation}
where $\theta_{i,n}\in[0,2\pi)$ denotes the phase shift of the $n$th reflecting element.

The interference term $I$ caused by other users is expressed as
\begin{equation}
    I = \sqrt{P} \left(\sum_{m}^{U_m}h_m+\sum_{i=1}^{R} v_i \boldsymbol{\Phi}_ih_{i} h_{mi}\right),
\end{equation}
where $U_I$ denotes the number of interfering users and $v_i$ is the ON/OFF state of RIS$_i$.

In dense, mobile scenarios, interference sources may include nearby or co-located users, onlookers, and smart wearable devices. Such interferences significantly degrade the quality of critical physiological data transmitted by target users in the uplink. Therefore, intelligent RIS control and user trajectory prediction are crucial for dynamically adjusting RIS states to mitigate interference~\cite{9090356}.

For the $l_{th}$ target user (denoted as U$_l$), the SINR at the BS can be expressed as
\begin{equation}
    \gamma_l = \frac{P \left| h_l+\sum_{i=1}^{R} v_i \boldsymbol{\Phi}_ih_{i} h_{li}\right|^2}
    {P \left| \sum_{m}^{U_m} h_m+\sum_{i=1}^{R} v_i \boldsymbol{\Phi}_ih_{i} h_{mi} \right|^2 + \sigma^2},
\end{equation}
where the numerator represents the effective signal power of user U$_l$ received at the BS, and the denominator consists of AWGN noise and interference signals.

Our goal is to ensure reliable data transmission to the target user by optimizing the RIS states and reflection coefficients to maximize the system SINR. 
Therefore, we introduce a binary activation control vector for the RISs, denoted by $\mathbf{V} = \{v_1, v_2, ..., v_R\}$, where $v_i \in \{0,1\}$ indicates the ON/OFF state of RIS$_i$. 
\section{Proposed Generative RIS Control Framework}\label{s4}

\begin{figure*}[h]
    \centering
    \includegraphics[width=0.95\linewidth]{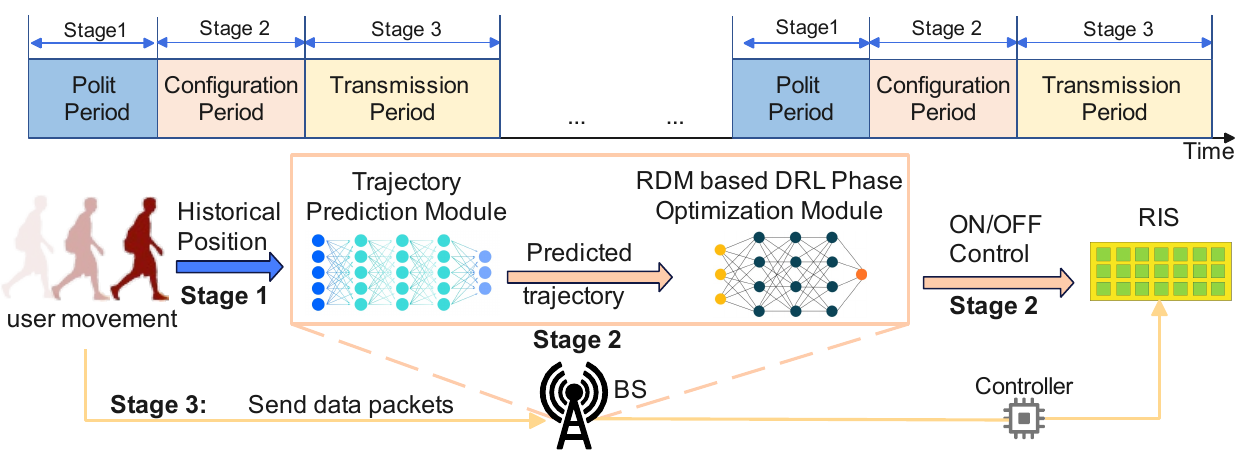} 
    \captionsetup{font={small}}
    \caption{The proposed TPGC structure. A trajectory prediction module forecasts future user locations, which are then used to predict the CSI of the cascaded channels. Conditioned on the predicted CSI, the proposed RDM-based DRL optimization module outputs the RIS ON/OFF decisions and passive beamforming to maximize the communication performance.}
    \label{fig:algorithm}
\end{figure*}
\subsection{Key idea}
Dynamic user mobility and the passive nature of RIS reflections fundamentally hinder communication reliability. Small geometric misalignments between moving users may inadvertently amplify co-channel interference when RIS panels blindly reflect all incident signals. To address this issue, we propose a TPGC framework that proactively configures RIS reflection behavior. It predicts users' future positions and generates geometry-consistent RIS phase configurations.

As illustrated in Fig.~\ref{fig:algorithm}, the TPGC framework comprises two synergistic modules. First, an LSTM-based trajectory-prediction module forecasts the future positions of target and interfering users. Second, an RDM with DRL guidance (RDM-DRL) optimizes the RIS phase directly on the torus manifold to ensure geometric consistency.
The final ON/OFF control strategy is determined by comparing the predicted achievable rates (ARs) for each RIS in both activated and deactivated states. This section provides the algorithmic motivations and methodological structure of both modules. Subsequent sections will present the mathematical formulations.

\subsection{Trajectory Prediction Module}

RIS configuration is particularly sensitive to user location. Users move constantly, so instantaneous Channel State Information (CSI) quickly becomes outdated. Relying solely on current CSI results in misaligned RIS phases and increased interference. To overcome these limitations, we employ a trajectory prediction module. This module uses a two-layer LSTM network augmented with dynamic motion descriptors, specifically speed and heading angle. This design is driven by the strong temporal dependencies in human running patterns and the need to predict geometric overlaps between users several steps ahead. The module outputs the predicted position of each user at time horizon $t$. The BS then uses this output to reconstruct the corresponding future BS-RIS-user cascaded CSI. This prediction-aware CSI serves as conditioning information for subsequent generative optimization.

\subsection{RDM-based DRL Phase Optimization Module}

Optimizing RIS phases in highly dynamic environments requires solving a high-dimensional, non-convex optimization problem on a periodic manifold. Existing DRL-based RIS optimizers rely on unimodal Gaussian actions and Euclidean updates \cite{du2024diffusion}. These methods fail to capture phase periodicity or to effectively explore multi-modal solution spaces. Similarly, prior diffusion-based optimizers in wireless systems operate exclusively in Euclidean space \cite{ni2025conditional}. This approach neglects the structure of RIS phase shifts and yields physically inconsistent solutions.

To address these issues, we propose an RDM that performs both noise injection and denoising directly on the torus manifold $(S^{1})^{N}$. This strategy ensures that every intermediate sample remains consistent with the phase-modulation geometry. Standard supervised diffusion training requires optimal labels, which are unavailable in RIS optimization. Therefore, our RDM incorporates DRL-guided refinement. We incorporate the AR gradient into the reverse diffusion process. This gradient guides the sampling toward optimal phase configurations. Consequently, this hybrid generative–reinforcement approach enables robust exploration under mobility. It avoids local optima and produces phase vectors that align well with predicted future CSI.

\subsection{RIS ON/OFF Control Module}

While the RDM-based DRL module produces continuous phase vectors under the assumption that each RIS operates in the ON state, activating all RIS panels indiscriminately may amplify interference in dense user scenarios. Therefore, after obtaining the optimized phase vector, the BS compares the predicted ARs with $\mathrm{RIS}_i$ switched ON and OFF. This enables the framework to suppress harmful reflections while preserving beneficial propagation links.

\subsection{Frame Structure Design}

The proposed TPGC is implemented in each transmission frame, which is divided into three phases.

\begin{itemize}
    \item \textbf{Stage~1: Pilot Period.} 
    The users transmit uplink pilot signals and positioning reference signals~\cite{ni2025conditional}. The BS estimates the instantaneous CSI for the direct and RIS-assisted links and extracts user location measurements (e.g., time-of-arrival, longitude, and latitude). These measurements are input to the LSTM-based trajectory prediction module to predict future user positions.
    
    \item \textbf{Stage~2: Configuration Period.} 
    The BS reconstructs the future CSI for horizon $t$ based on the predicted positions, as detailed in Section~\ref{s3}. Using $\widehat{\mathbf{H}}(t)$ as the conditioning information, the RDM generates candidate RIS phase configurations following Algorithm~\ref{alg:RDM_phase}. The DRL-guided refinement steers the diffusion trajectory toward high-achievable-rate phase vectors. After quantization, the AR under RIS ON and OFF are compared, and the final phase vector $\boldsymbol{\Phi}^{\star}$ and ON/OFF state $v^{\star}$ are determined.
    
    \item \textbf{Stage~3: Transmission Period.} 
    In this phase, the RIS implements the optimized configuration by activating only the panels for which $v^{\star}=1$ and applying the corresponding phase vector $\boldsymbol{\Phi}^{\star}$. Users transmit uplink payload data, and the optimized RIS reinforces the desired signal while suppressing interference, thereby improving end-to-end link quality and overall system throughput.
\end{itemize}

\section{LSTM Trajectory Prediction Mechanism}\label{s5}
According to 3GPP Release 18 \cite{R18} and Release 19 \cite{R19}, the BS can obtain and manage user location information with an accuracy ranging from meter-level to tens of meters. With the forthcoming deployment of 6G networks, intelligent models can be deployed at the BS to enable real-time location prediction while minimizing signaling overhead compared to conventional methods.

In this study, we use real-world trajectory data collected in Beijing, drawn from the comprehensive GeoLife trajectory dataset \cite{geolife}, which provides rich spatio-temporal data for collaborative social networking services. The dataset includes user longitude and latitude over time, enabling detailed analysis of human mobility patterns \cite{pappalardo2015returners}. This model can predict both the positions of the primary user and the potential interfering users, thereby providing a foundation for subsequent RIS configuration\footnote{To support reproducibility, the processed trajectory traces used in our simulations are publicly available at \url{https://github.com/wangxn2/TPGC}.}.

\subsection{Trajectory Representation and Prediction Features}

Let $\mathbf{q}_u(t) = \big[q_{u,1}(t),\, q_{u,2}(t)\big]^\top \in \mathbb{R}^2$ denote the location of $U_u$ at the time index $t$, where $q_{u,1}(t)$ and $q_{u,2}(t)$ are the two planar coordinates. A conventional LSTM predictor uses a length-$H$ history window, i.e.,
$\mathcal{Q}_u(t) = \{\mathbf{q}_u(t-H+1),\ldots,\mathbf{q}_u(t)\}$ as input and outputs the next-step location estimate,  given by
$
\widehat{\mathbf{q}}_u(t) = f_{\psi}\!\big(\mathcal{Q}_u(t-1)\big)$,
where $f_{\psi}(\cdot)$ denotes an LSTM network parameterized by $\psi$. This coordinate-only baseline captures temporal correlations in the position sequence but may become less robust when users exhibit non-uniform motion, such as acceleration, deceleration, or turning~\cite{mobility}.

We augment the coordinate sequence with two kinematic features computed from consecutive locations. Specifically, we define the displacement magnitude between two successive samples as
\begin{equation}
\Delta d_u(t) = \big\|\mathbf{q}_u(t)-\mathbf{q}_u(t-1)\big\|,
\end{equation}
and let $\Delta \tau$ be the sampling interval. The instantaneous speed is computed as
\begin{equation}
\nu_u(t) = \frac{\Delta d_u(t)}{\Delta \tau}.
\end{equation}
The motion direction is obtained from the displacement vector
\begin{equation}
\vartheta_u(t) \!=\! \mathrm{atan2}\!\Big(q_{u,2}(t)\!-\!q_{u,2}(t\!-\!1),\, q_{u,1}(t)\!-\!q_{u,1}(t\!-\!1)\Big),
\end{equation}
where $\mathrm{atan2}(\cdot,\cdot)$ denotes the four-quadrant inverse tangent function that returns the orientation angle of the displacement vector.
With these features, the input at time $t$ is
$
\boldsymbol{\zeta}_u(t)=\big(\mathbf{q}_u(t),\, \nu_u(t),\, \vartheta_u(t)\big).
$

\subsection{Offline Training and Online Inference}
We implement the trajectory predictor $f_{\psi}(\cdot)$ using an LSTM. This architecture is well suited to modeling temporal dependencies in sequential mobility data~\cite{mobility}. The model architecture consists of two LSTM layers to capture long-term temporal dependencies. After LSTM encoding, a fully connected layer maps the hidden state to the predicted coordinates for the next time step. This structure includes a memory update mechanism that adjusts its output based on current inputs and historical states, which enhances temporal modeling accuracy.
We train the model by minimizing the mean squared error (MSE) using the Adam optimizer \cite{Croitoru2023}.

After training, we deploy the LSTM predictor at the BS for real-time operation. During online inference, the BS continuously stores the most recent $H$ observations for each user. It constructs the associated feature sequence, and computes $\widehat{\mathbf{q}}_u(t)$ at each control step. The predicted location is subsequently passed to the geometric channel reconstruction module, which generates the CSI for RIS phase configuration and ON/OFF control.

\subsection{Problem Analysis}
At each control step $t$, the BS initially retrieves the predicted locations $\widehat{\mathbf{q}}_u(t)$ of all users from the trajectory prediction module. Since a deterministic LoS geometric channel model is used in Section~\ref{s3}, these predicted locations uniquely determine the corresponding propagation distances, thereby enabling CSI reconstruction. In particular, for each user, the predicted location $\widehat{\mathbf{q}}_u(t)$ is mapped directly to the BS-user distance $d_u(t)$ and the user-RIS distance $d_{ui}(t)$, which form the foundation for reconstructing the direct and RIS-assisted channels.

Accordingly, the predicted direct BS-user channel of user $U_u$ at time $t$ is expressed as
\begin{equation}
    h_{u}(t)=d_u(t)^{-\alpha/2}
    exp\left(-j\frac{2\pi }{\lambda}d_u(t)\right),
\end{equation}
where $d_u(t)$ denotes the distance between the BS and user $U_u$ computed from $\widehat{\mathbf{q}}_u(t)$.
Similarly, the predicted user-RIS channel is given by
\begin{equation}
    h_{ui}(t) =
    \frac{1}{d_{ui}(t)^{\alpha/2}}
    \exp\!\left(-j\frac{2\pi}{\lambda} d_{ui}(t)\right).
\end{equation}

Based on the predicted user locations, the BS reconstructs the predicted
CSI at time $t$ as
\begin{equation}
\begin{aligned}
    \hat{\mathbf H}(t)=
\Big\{\hat h_l(t),\,\hat h_{m}(t),\,\hat h_{li}(t),\,
\hat h_{mi}(t),\,h_i\;\big|\;&
l\in\mathcal U_l,m\in\mathcal U_m, \\
&i\in\mathcal R\Big\}.
\end{aligned}
\end{equation}

Based on the reconstructed CSI, the predicted SINR of the target user $U_l$ at time $t$ can be directly written as
\begin{equation}\label{gammal}
\gamma_l(t)=\frac{P \left|h_l(t)+\sum\limits_{i=1}^{R} v_i\boldsymbol{\Phi}_i(t) h_i(t) h_{ui}(t)\right|^2
}{P \left|\sum\limits_{m=1}^{U_I} h_m(t)+\sum\limits_{i=1}^{R} v_i\boldsymbol{\Phi}_i(t) h_i(t) h_{mi}(t)\right|^2
+\sigma^2}.
\end{equation}

Using the predicted CSI reconstructed from $\widehat{\mathbf{q}}_u(t)$, the RIS phase and ON/OFF configuration problem at time $t$ is reformulated as
\begin{equation}
\begin{aligned}
\mathcal{P}_1:\quad \max_{\mathbf{V}(t),\,\boldsymbol{\Phi}(t)}\quad &\sum_{l=1}^{L}
 \log_2(1+\gamma_l(t))\\
\text{s.t.}\quad
& v_i(t)\in\{0,1\},\quad \forall i \in \mathcal{R},\\
& |e^{j\theta_n^i(t)}| \!=\! 1, \forall i \in \mathcal{R},\ \forall n \in \{1,N\}.
\end{aligned}
\label{problem:PTPC}
\end{equation}

Problem~\eqref{problem:PTPC} is a mixed-integer, non-convex optimization problem defined on the torus manifold of RIS phase shifts. is classified as a mixed-integer nonlinear programming (MINLP) problem. The objective or constraints include at least one nonlinear term, and the optimization variables comprise both discrete (e.g., $v_i(t)$) and continuous (e.g., $\theta(t)$) variables, resulting in a hybrid discrete-continuous decision space typical of such problems. In the next section, we propose an RDM with DRL guidance to solve this trajectory-predicted control problem effectively.

\section{Generative ON/OFF Control Mechanism}\label{s6}

\begin{figure*}[h]
    \centering
    \includegraphics[width=0.96\linewidth]{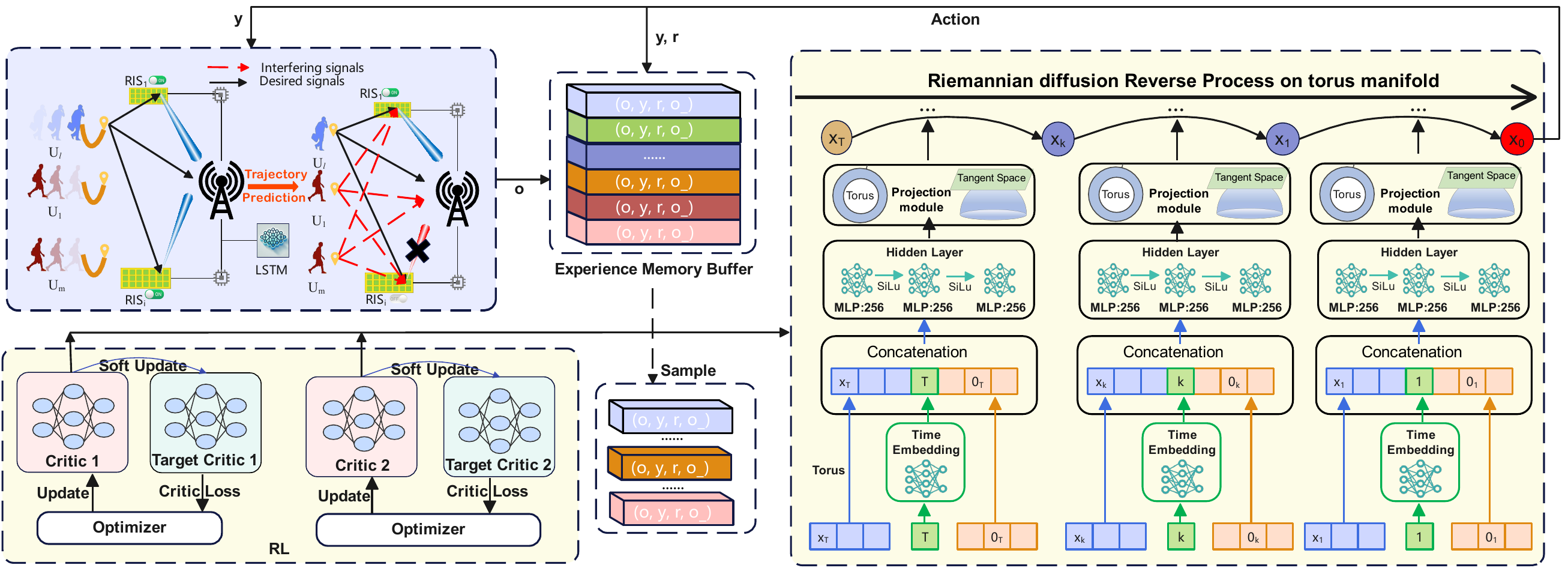}
    \captionsetup{font={small}}
    \caption{Architecture of the proposed RDM–DRL framework for trajectory-predicted RIS phase and ON/OFF control. The framework takes the predicted cascaded CSI, reconstructed from future user trajectories, as the conditioning observation. The RDM is employed as a generative actor that samples RIS phase configurations directly on the torus manifold. The reverse diffusion process is guided by a TD3-based critic, which evaluates the achievable rate and steers the denoising trajectory toward high-quality phase solutions. After obtaining the optimized continuous phase configuration, the achievable rates under RIS activation and deactivation are compared to determine the final ON/OFF control decision.}
    \label{fig:RDMRL}
\end{figure*}
After predicting future user positions and reconstructing the corresponding cascaded CSI at time $t$, the problem (\ref{problem:PTPC}) can be solved using a two-stage control strategy. Specifically, we first optimize the continuous RIS phase configuration under the assumption that the RIS is activated, and then determine the binary ON/OFF control state by comparing the AR under the RIS-ON and RIS-OFF modes.

\begin{algorithm}[t]
\caption{Generative ON/OFF Control Mechanism}
\label{alg:RDM_phase}
\begin{algorithmic}[1]
\State \textbf{Input:} Predicted CSI $\widehat{\mathbf{H}}(t)$
\State \textbf{Output:} Optimized phase vector $\boldsymbol{\Phi}^{\star}$ and ON/OFF state $v^{\star}$

\State Construct observation $\mathbf{o}$ from $\widehat{\mathbf{H}}(t)$
\State Initialize latent variable $\mathbf{x}_T$ on the torus manifold $(S^1)^N$

\For{$k = T, T-1, \ldots, 1$}
    \State Compute denoising output $\widehat{\boldsymbol{\epsilon}}_k$ using (\ref{eq:denose})
    \State Perform reverse diffusion update according to (\ref{eq:update})
    \If{$k>1$}
        \State Sample tangent noise and update $\mathbf{x}_{k-1}$ via (\ref{eq:reverse_update})
    \Else
        \State Project $\mathbf{x}_{k-1}$ onto the torus using (\ref{eq:reverse_update})
    \EndIf
\EndFor

\State Recover continuous phase vector $\boldsymbol{\Phi}^{\star}$ using~(\ref{eq:theta_recovery})
\State Compute achievable rates $AR_{\rm on}(t;\boldsymbol{\Phi}^{\star})$ and $AR_{\rm off}(t)$
\State Determine ON--OFF decision $v^{\star}$ according to (\ref{eq:onoff_decision})
\end{algorithmic}
\end{algorithm}

\subsection{RIS Phase Optimization via Riemannian Diffusion}
The RIS phase optimization problem is inherently high-dimensional and non-convex, subject to strong periodic constraints. Each reflecting element introduces a unit-modulus phase shift, and the complete phase vector lies on the torus manifold $(S^{1})^{N}$, with $N$ representing the number of RIS elements. Recent generative DRL studies \cite{zhang2024multi} highlight a critical limitation in current approaches. Conventional Euclidean actor networks \cite{liang2025cross} often fail to capture the complex distribution of optimal actions within high-dimensional spaces. In the context of RIS, ignoring the intrinsic geometry may lead to discontinuities at the $ 0$–$2\pi$ boundary. It also yields intermediate solutions that violate the unit-modulus constraint.

To address this issue, we utilize an RDM \cite{RDM} to construct a geometry-aware generative actor model. This model constructs a geometry-aware generative actor. Our core philosophy formulates phase optimization as a conditional generation task directly on the manifold. We do not define diffusion directly in terms of phase angles. Instead, we employ an extrinsic embedding of the torus manifold in Euclidean space. This approach enables efficient implementation while rigorously preserving the manifold structure. Moreover, the diffusion model is integrated within a twin delayed deep deterministic policy gradient (TD3) framework to overcome the lack of optimal labels. Here, the diffusion process serves as the expressive policy $\pi_{\phi}$. Meanwhile, the twin critics implicitly guide the denoising trajectory toward high-reward regions. This combination yields an RL-guided generative-phase optimizer that remains robust to CSI uncertainties.

\subsection{Riemannian Diffusion Formulation on the Torus}

Let $\boldsymbol{\Phi}=[\theta_1,\ldots,\theta_N]^{\mathsf{T}}$ denote the RIS phase vector.
We represent each phase $\theta_n$ by a 2D unit vector and embed the torus manifold $(S^1)^N$ into $\mathbb{R}^{2N}$ as
\begin{equation}
\mathbf{y}=\big[(\cos\theta_1,\sin\theta_1),\ldots,(\cos\theta_N,\sin\theta_N)\big]\in \mathbb{R}^{2N},
\label{eq:torus_embedding}
\end{equation}
where each 2D block corresponds to one RIS element.
This extrinsic representation follows the Stratonovich formulation of Riemannian diffusion \cite{StochasticIntegration}. It allows us to implement diffusion dynamics in the ambient space while respecting the manifold geometry \cite{RDM}.

Let us denote by $\mathbf{x}_k\in\mathbb{R}^{2N}$ the latent variable at diffusion step $k$.
The projection onto the torus manifold is performed blockwise as
\begin{equation}
\Pi_{\mathcal{T}}(\mathbf{x})
=
\left[
\frac{\mathbf{x}_1}{\|\mathbf{x}_1\|_2},\ldots,
\frac{\mathbf{x}_N}{\|\mathbf{x}_N\|_2}
\right],
\label{eq:proj_torus}
\end{equation}
where $\mathbf{x}_n\in\mathbb{R}^2$ denotes the $n$-th block.

To ensure stochastic perturbations remain on the tangent space, the tangent projection is given by
\begin{equation}
\Pi_{T_{\mathbf{x}}\mathcal{T}}(\boldsymbol{\xi})
=
\big[
\boldsymbol{\xi}_n-(\mathbf{x}_n^{\mathsf{T}}\boldsymbol{\xi}_n)\mathbf{x}_n
\big]_{n=1}^{N}.
\label{eq:proj_tangent}
\end{equation}

Injecting stochastic perturbations in the tangent space is essential to preserve the intrinsic geometry of the torus manifold~\cite{RDM}. Under the Stratonovich interpretation~\cite{StochasticIntegration}, tangent-space noise guarantees that the diffusion trajectory remains on the manifold without introducing spurious curvature-induced drift terms.

We employ a $T$-step diffusion process with a predefined variance schedule $\{\beta_k\}_{k=1}^{T}$, where $\alpha_k=1-\beta_k$ and $\bar{\alpha}_k=\prod_{z=1}^{k}\alpha_z$.
Given the observation $\mathbf{o}$ constructed from the predicted CSI $\widehat{\mathbf{H}}(t)$, yields:
\begin{equation}
o(t)=\hat{\mathbf H}(t).
\end{equation}
The denoising network outputs are then given by:
\begin{equation}\label{eq:denose}
\widehat{\boldsymbol{\epsilon}}_k
=
\varepsilon_{\phi}(\mathbf{o},\mathbf{x}_k,k),
\end{equation}
which defines the reverse diffusion update in the ambient space as 
\begin{equation}\label{eq:update}
\mathbf{G}_k
=\frac{1}{\sqrt{\alpha_k}}\left(\mathbf{x}_k-\frac{\beta_k}{\sqrt{1-\bar{\alpha}_k}}\widehat{\boldsymbol{\epsilon}}_k\right).
\end{equation}
The update $\mathbf{G}_k$ represents a critic-guided drift prediction
in the ambient space. Unlike conventional diffusion models, the parameters
$\{\alpha_k,\beta_k\}$ do not correspond to a forward noising process,
but instead, control the step size and stochasticity of the guided
reverse diffusion trajectory.

For $k>1$, a stochastic tangent perturbation is applied:
\begin{equation}
\mathbf{x}_{k-1}= \Pi_{\mathcal{T}} \left(\mathbf{G}_k+ \sqrt{\tilde{\beta}_k}\, \Pi_{T_{\mathbf{x}_k}\mathcal{T}} (\boldsymbol{\xi}_k) \right), \quad \boldsymbol{\xi}_k\sim\mathcal{N} (\mathbf{0},\mathbf{I}),
\label{eq:reverse_update}
\end{equation}
where $\tilde{\beta}_k$ controls the magnitude of the stochastic
tangent perturbation and serves as an exploration variance in the
guided diffusion process.

After completing all reverse steps, the continuous phase vector is recovered by
\begin{equation}
\theta_n=\mathrm{atan2}\big([\mathbf{x}_{0,n}]_2,[\mathbf{x}_{0,n}]_1\big),
\quad n=1,\ldots,N.
\label{eq:theta_recovery}
\end{equation}

\subsection{RL-Guided Optimization Objective}

The above diffusion sampler defines a stochastic policy
$\pi_{\phi}(\boldsymbol{\Phi}\mid\mathbf{o})$.
The reward is defined as the AR of the desired user under RIS activation:
\begin{equation}
r(\mathbf{o},\boldsymbol{\Phi})
=
AR_{\rm on}(t;\boldsymbol{\Phi})
=
\log_2\big(1+\gamma_l(t)\big),
\end{equation}
where $\gamma_l(t)$ is computed using equation ~\ref{gammal}.

We adopt the TD3 framework to guide the diffusion actor.
The twin critics are trained via the Bellman equation\cite{TD3}, while the actor parameters $\phi$ are optimized by
\begin{equation}
\max_{\phi}
\;
\mathbb{E}_{\mathbf{o}}
\Big[
Q_{\psi_1}\big(\mathbf{o},\pi_{\phi}(\mathbf{o})\big)
\Big].
\end{equation}
The critic gradients are backpropagated through the unrolled reverse diffusion steps, thereby implicitly refining the denoising network and steering the Riemannian diffusion trajectory toward high-rate regions on the torus manifold. Specifically, the reverse diffusion process induces a differentiable
mapping $\boldsymbol{\Phi}=g_{\phi}(\mathbf{o},\boldsymbol{\xi})$,
where $\boldsymbol{\xi}$ collects all stochastic perturbations.
This enables end-to-end policy optimization via the chain rule.

\subsection{ON/OFF Control Decision}

The optimization problem in~\eqref{problem:PTPC} involves both continuous phases and binary ON/OFF states. In our framework, Riemannian diffusion and RL guidance are first employed to obtain a high-quality continuous-phase solution, assuming the RIS is activated.
Let $\boldsymbol{\Phi}^{\star}$ denote the final continuous phase vector generated by the diffusion actor.
The AR with RIS activated is given by
\begin{equation}
AR_{\rm on}^{\star}
=
AR_{\rm on}(t,\boldsymbol{\Phi}^{\star}),
\end{equation}
while the AR with RIS deactivated is obtained from the direct link as
\begin{equation}
AR_{\text{off}}(t)=\log_2\!\left(1+\gamma_l(t)\big|_{v_i=0,\forall i}\right).
\end{equation}
The ON/OFF control decision is then obtained as 
\begin{equation}
v^{\star}
=
\begin{cases}
1, & AR_{\rm on}^{\star} \ge AR_{\rm off}(t),\\
0, & \text{otherwise}.
\end{cases}
\label{eq:onoff_decision}
\end{equation}

\subsection{Main Flow of Proposed Algorithm}
At the start of each control interval $t$, the BS reconstructs the predicted cascaded CSI $\widehat{\mathbf{H}}(t)$ from the estimated user locations. This information is mapped to the observation vector $\mathbf{o}(t)$, which encodes the effective direct and RIS-assisted channels, the interference structure, and the system configuration parameters. The observation serves as the conditioning input to both the diffusion-based actor and critic networks. Conditioned on $\mathbf{o}(t)$, the Riemannian diffusion actor initializes a latent variable $\mathbf{x}_T$ on the torus manifold $(S^1)^N$. Through $T$ reverse diffusion steps, the model iteratively denoises $\mathbf{x}_k$ according to (\ref{eq:denose})-(\ref{eq:reverse_update}), where each update consists of denoising in the ambient space, tangent-space stochastic perturbation, and projection back to the torus manifold. Unlike supervised diffusion models, the proposed RDM does not rely on optimal phase labels. Instead, the twin TD3 critics evaluate the achievable rate $AR_{\rm on}(t;\boldsymbol{\Phi})$ resulting from the generated phase vector. The critic gradients are backpropagated through the unrolled reverse diffusion steps, refining the denoising network parameters and guiding the sampling trajectory toward high-reward regions. Upon completion of the reverse diffusion process, the final latent variable $\mathbf{x}_0$ is mapped to the continuous phase vector $\boldsymbol{\Phi}^{\star}$ using (\ref{eq:theta_recovery}). The corresponding achievable rate under RIS activation, $AR_{\rm on}^{\star}$, is computed using the predicted SINR expression in (\ref{gammal}). To assess whether RIS activation is beneficial under the predicted interference conditions, the BS compares $AR_{\rm on}^{\star}$ with the achievable rate under RIS deactivation, $AR_{\rm off}(t)$. The final ON/OFF decision $v^{\star}$ is made according to (\ref{eq:onoff_decision}), suppressing harmful reflections while preserving constructive signal enhancement.

\subsection{Complexity Analysis}
The dominant cost of the RDM actor stems from the reverse diffusion process.
At each denoising step, the network processes a $2N$-dimensional embedded phase vector, which results in a complexity of $\mathcal{O}(N w_{\rm net})$, where $w_{\rm net}$ denotes the width of the denoising network.
After $T$ reverse steps, the overall diffusion sampling complexity is $\mathcal{O}\big(T N w_{\rm net}\big)$\cite{du2024diffusion}. 
The projection and tangent-space operations in (\ref{eq:proj_torus}) and (\ref{eq:proj_tangent}) introduce a linear overhead $\mathcal{O}(N)$ and are thus negligible in comparison to neural inference. 
The twin TD3 critics evaluate the achievable rate based on the generated phase vector, with per-evaluation complexity of $\mathcal{O}(N)$.
During training, critic backpropagation through unrolled diffusion steps adds a factor proportional to $T$, resulting in $\mathcal{O}(T N)$ time complexity~\cite{TD3}.
In online inference, however, only forward diffusion sampling and one ON/OFF comparison are needed. 
The ON/OFF control requires computing $AR_{\rm on}(t;\boldsymbol{\Phi}^{\star})$ and $AR_{\rm off}(t)$, both of which scale linearly with the number of RIS elements $N$ and users $U$, resulting in $\mathcal{O}(UN)$ time complexity.

Combining the above components, the total online time complexity per control interval is $\mathcal{O}\left(TNw_{\rm net}+UN\right)$, which scales linearly with $N$ and $T$~\cite{DDRL}. Since $T$ is small in practice (e.g., $T=4$ in simulations), the proposed method remains computationally tractable for large-scale RIS deployments. Moreover, the diffusion sampling and critic evaluations are amenable to parallelization on GPUs, making the proposed RDM-DRL framework well-suited for real-time implementation at the BS.

\section{Simulation Results and Discussion} \label{s7}
\subsection{Simulation Setup}
\subsubsection{Simulation Platform}
The experiments were conducted on a workstation equipped with an NVIDIA GeForce RTX 4090 GPU (24 GB), an Intel(R) Xeon(R) Silver 4410Y CPU, and 128 GB of RAM, running Ubuntu 22.04.1 LTS. All deep learning models were implemented in PyTorch 2.5.1 with CUDA 12.2 support. To support reproducibility, the processed trajectory traces used in our simulations are publicly available at \url{https://github.com/wangxn2/TPGC}.

\subsubsection{Implementation for Trajectory Prediction}
To comprehensively evaluate the performance of the proposed trajectory prediction model, we use the real-world GeoLife trajectory dataset \cite{geolife} and select multiple representative user trajectory files as test samples. The proposed trajectory prediction framework is built on an LSTM architecture. The input consists of historical trajectory samples (longitude, latitude, speed, and heading angle) over consecutive time steps, and the output is the predicted future trajectory coordinates.

During training, we select 71 trajectories from the user trajectory data collected in Beijing for training and evaluation. A sliding-window approach is used to construct input sequences, with each sequence comprising five consecutive trajectory points to improve model generalization. The model employs a two-layer LSTM that captures time-series features by retaining historical information and updating the hidden state. For the loss function, we adopt the MSE as the training objective.

During inference, given user trajectory data, the trained LSTM model predicts future trajectory points. To align with position-accuracy requirements in wireless communication standards, we use the Haversine formula to compute the geodesic distance between predicted and ground-truth coordinates, and report the mean geodesic error (in meters) as the primary metric~\cite{meters}, thereby quantifying the spatial accuracy of trajectory prediction.

\subsubsection{Simulation settings}
Unless otherwise specified, the simulation model consists of one BS, $10$ RISs, and $10$ interfering users. Each RIS contains $600$ reflecting elements, and the RISs are uniformly placed on a circle of radius $10$ m centered at the BS. We assume that the RIS reflection efficiency decreases with the angle of incidence measured from the surface normal and is maximized at normal (perpendicular) incidence. The simulation setup aligns with common practices in RIS system-level evaluation and accounts for distributed RIS deployment to enhance coverage. The target user's transmit power is set to $1$ W, and the noise power is $1$ pW.

\subsubsection{Model Design}
A conditional network is composed of two self-attention blocks with 256 hidden dimensions, sinusoidal time embeddings, and a fully connected output head, which predicts the torus-manifold function for each RIS element. The diffusion process uses four denoising steps with a cosine step-size schedule, a von Mises forward-noise process with a quadratically decayed concentration parameter, and a guidance coefficient $\eta=0.1$ for reward-gradient-based refinement. All phases are represented on $(S^{1})^{N}$ and updated modulo $2\pi$. The model is trained with Adam using a learning rate of $10^{-4}$, a batch size of 128.

\subsubsection{Benchmarks}
To comprehensively evaluate the proposed RDM–DRL optimizer, we compare it with six representative baseline models widely adopted in RIS and wireless control tasks:

\begin{itemize}
\item \textbf{Euclidean Diffusion with DRL (EDD)} \cite{ho2020denoising}:
A standard DDPM constructed in Euclidean space with Gaussian forward noise and UNet-style denoiser, ignoring the phase-manifold geometry.
\item \textbf{DDPG \cite{DDPG}:}
An actor–critic model with deterministic Gaussian policy, suited for continuous RIS phase control but prone to local minima under high-dimensional action spaces.
\item \textbf{TD3 \cite{TD3}:}
A twin-critic DDPG variant with delayed updates to improve stability; serves as a strong continuous-control baseline.
\item \textbf{SAC \cite{SAC}:}
A stochastic maximum-entropy policy model generating diverse actions, improving exploration for RIS optimization.
\item \textbf{PPO \cite{PPO}:}
A clipped-policy gradient method providing stable training for moderate-dimensional RIS phase control.
\item \textbf{A3C \cite{A3C}:}
An asynchronous advantage actor–critic method that leverages parallel agents to stabilize policy updates, offering improved robustness and faster convergence in dynamic RIS control scenarios.
\end{itemize}
All the learning-based baselines share identical definitions of state, action, and reward for fairness.

We additionally benchmark the proposed control strategy against four RIS-configuration policies:
\begin{itemize}
\item \textbf{TPC (Trajectory-Predicted Control):}
The baseline method combines LSTM-based trajectory prediction with geometric CSI reconstruction and ON/OFF decision.
\item \textbf{RIS Always-On:}
A naive configuration where all RIS panels remain activated, representing the upper bound of pure signal enhancement but suffering from severe interference amplification.
\item \textbf{ISL-Based Control \cite{tvt}:}
An intelligent spectrum learning scheme that classifies desired and interfering signals using CNN-based features and toggles RIS ON/OFF accordingly, but without future trajectory foresight.
\end{itemize}
\subsection{Trajectory Prediction}

\begin{figure}[t]
    \centering
    \captionsetup{font={small}}
     \includegraphics[width=1\columnwidth]{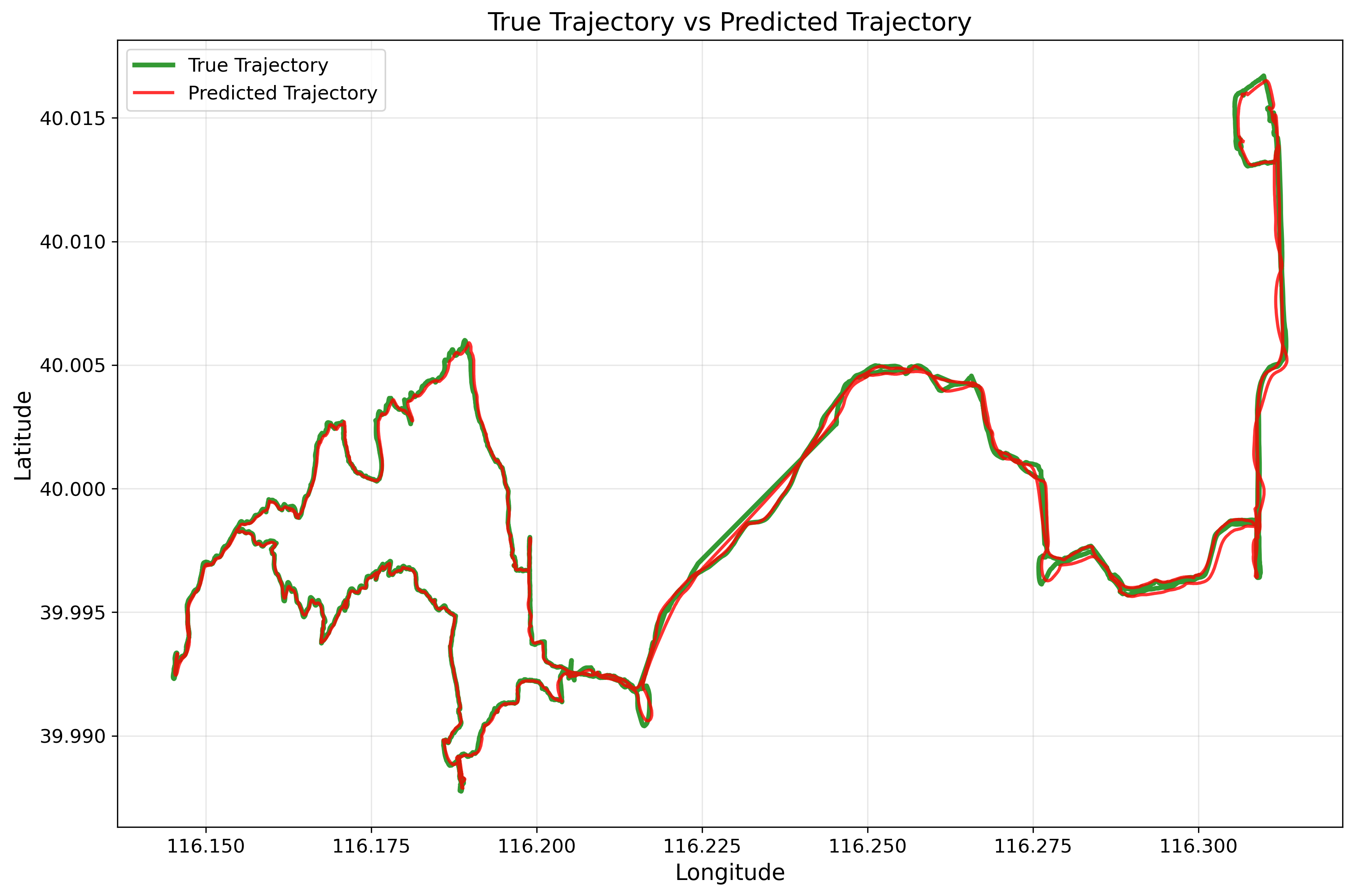}
    \caption{Comparison of true trajectory with predicted trajectories.}
    \label{fig:result3}
\end{figure}

To evaluate the effectiveness of the trajectory prediction model in realistic competition scenarios, we selected a representative athlete’s trajectory as a test sample. We used the trained LSTM to predict future positions along it. As shown in Fig.~\ref{fig:result3}, the green solid line represents the ground-truth trajectory of the athlete, and the red dashed line denotes the predicted trajectory. Visual inspection indicates that the predicted trajectory closely matches the ground-truth trajectory.

At key turning points, long-distance straight segments, and regions with abrupt trajectory changes, the predicted curve continues to track the athlete’s motion direction, demonstrating strong spatiotemporal modeling capability. The experimental results indicate that the prediction errors of most test trajectories remain within $\pm 6$ m.

This result demonstrates that the proposed LSTM model can capture the overall evolution of the athlete’s trajectory while remaining robust to abrupt changes in local segments. In dense mobile scenarios, this model can serve as a key module to provide high-quality trajectory information for task scheduling in downstream intelligent communication systems, particularly for dynamic RIS ON/OFF control, resource coordination, and interference avoidance.

\subsection{Generalization}

\begin{figure}[t]
    \centering
    \captionsetup{font={small}}
    \includegraphics[width=1\columnwidth]{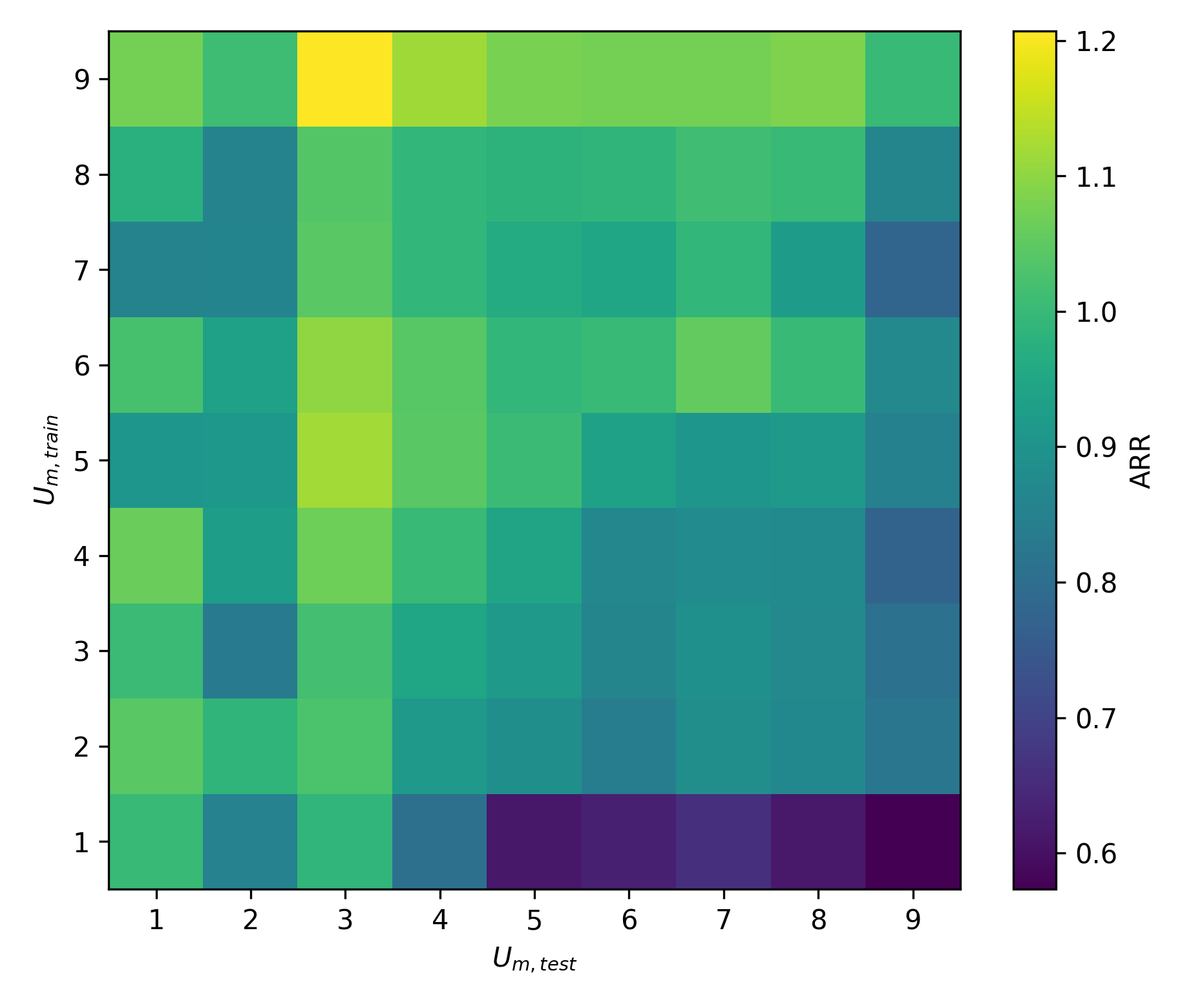}
    \caption{ARR matrix under cross-scenario interference mismatch, where models trained with different numbers of interfering users are evaluated across varying deployment interference densities.}
    \label{fig:Matrixconfusion}
\end{figure}

To evaluate the robustness and generalization capability of the proposed TPGC framework under environmental mismatch, we conduct a cross-scenario generalization study across varying interference densities. Let $m_{\mathrm{train}}$ denote the number of interfering users considered during training, and let $m_{\mathrm{test}}$ denote the actual interference density in the deployment environment.

For each interference scenario indexed by $m$, we define the {achievable rate ratio} (ARR) as
\begin{equation}
\mathrm{ARR}_m
=
\frac{\mathrm{AR}_{m}^{\mathrm{test}}}
     {\mathrm{AR}_{m}^{\mathrm{train}}}
\label{eq:reward_ratio}
\end{equation}
where $\mathrm{AR}_{m}^{\mathrm{train}}$ denotes the average achievable rate obtained by the trained model under its corresponding training interference configuration, and $\mathrm{AR}_{m}^{\mathrm{test}}$ denotes the achievable rate achieved by the same model when evaluated in an environment with $U_{m}^{\mathrm{test}}$ interfering users.

An ARR close to $1$ indicates that the trained policy maintains consistent performance under interference mismatch. In contrast, values greater than $1$ imply favorable scene migration and improved performance when deployed in environments less challenging than the training scenario.

Fig.~\ref{fig:Matrixconfusion} shows the ARR for $U_{m}^{test}, U_{m}^{train} \in {1,\ldots,9}$. The diagonal entries remain close to unity, indicating that the framework maintains high performance when the training and deployment interference conditions are aligned. More importantly, when $U_{m}^{train} > U_{m}^{test}$, several off-diagonal regions show reward ratios greater than $1$, indicating that models trained under denser interference conditions generalize effectively to environments with fewer interferers.

This phenomenon suggests that the RDM-DRL architecture learns interference-robust structural representations of RIS-assisted channels rather than overfitting to a particular interference level. Overall, the results demonstrate the robust generalization capability of the proposed framework across diverse multi-user interference configurations.

\subsection{Comparison with Benchmarks}
\begin{figure}[t]
    \centering
    \captionsetup{font={small}}
    \includegraphics[width=1\columnwidth]{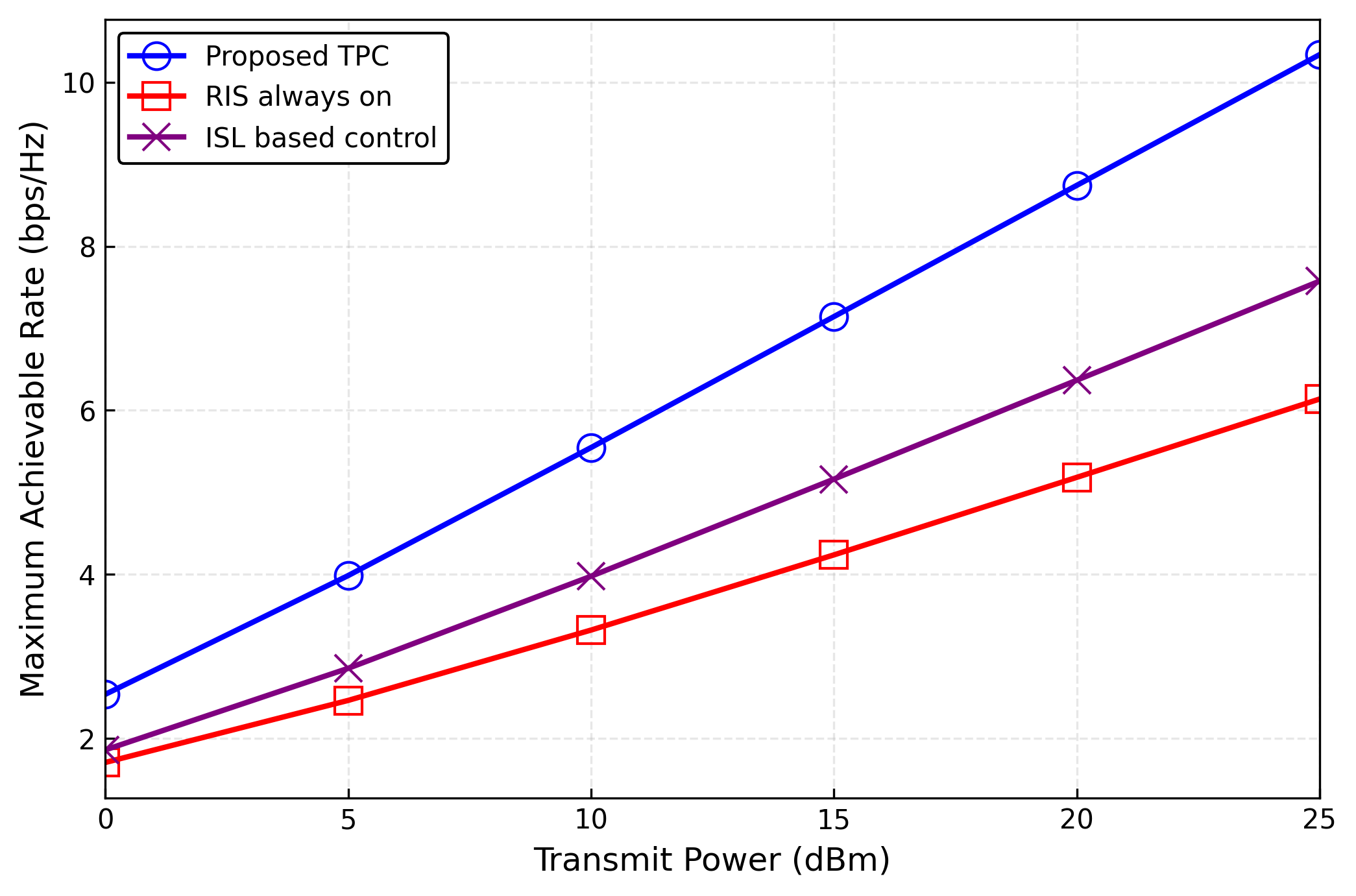}
    \caption{SINR performance analysis across different transmission power levels comparing TPC, RIS Always On, and ISL control methods.}
    \label{fig:power_analysis}
\end{figure}
To clearly demonstrate the advantages of the proposed method, this study compares the proposed TPGC algorithm with two baseline methods under different transmission powers and RIS configurations with varying numbers of elements. One is the ``RIS Always On" mode, and the second is the ISL control strategy based on prior work \cite{tvt}.

In the competition scenario, we first select a historical trajectory from the Geolife trajectory dataset, use an LSTM model to predict its path over the next $50$ seconds, and concurrently predict the trajectories of $10$ interfering users. Then, based on the prediction results and the communication model, we calculate, compare, and analyze the target user's SINR and evaluate its variation under different transmission power and RIS count conditions.

As shown in Fig.~\ref{fig:power_analysis}, increasing transmission power increases the SINR for all three methods. However, the proposed TPC method can predict the user's future position, identify the interference source, and dynamically control the ON/OFF states of each RIS, enabling real-time interference avoidance and thus achieving the highest SINR. In contrast, the ISL control method fails to predict future interference. It responds solely to the current state, making it difficult to adapt to dynamic changes in the trajectory scenario. Additionally, the ``RIS Always On" scheme fails to identify the interference source and instead amplifies the interference signal in areas of strong interference, performing the worst. Particularly in competitive environments with high interference density, blind reflection can cause interference power to exceed the desired signal, significantly degrading communication quality.

\subsection{Comparison with other Models}
\begin{figure}[t]
    \centering
    \includegraphics[width=1\columnwidth]{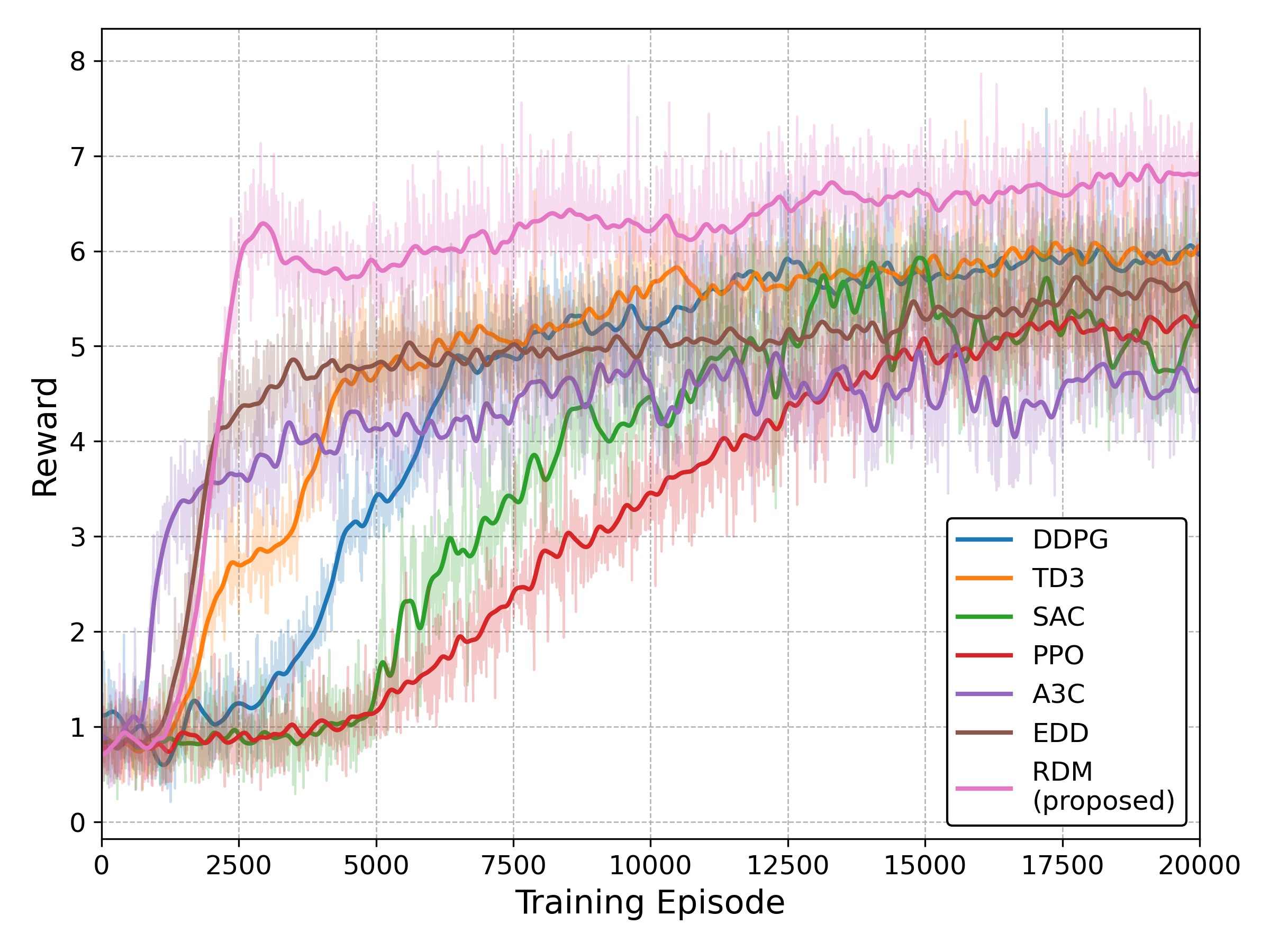}
    \caption{Reward versus training steps for different learning-based optimization methods.}
    \label{fig:models}
\end{figure}

To evaluate the effectiveness of the proposed framework, we compare seven representative learning models under identical RIS-assisted uplink configurations, e.g., DDPG, TD3, PPO, SAC, A3C, EDD, and the proposed RDM. All models are evaluated under the same state representation, reward formulation, and interaction environment to ensure a fair comparison. Each model is trained for $2 \times 10^{4}$ epochs, intentionally chosen to fully capture the convergence efficiency of all baseline methods, some of which exhibit noticeably slower convergence rates in high-dimensional RIS phase optimization problems. The average episodic reward is tracked to assess learning stability and convergence behavior.

Fig.~\ref{fig:models} plots the learning curves. All methods eventually converge, thereby confirming the learnability of the RIS phase optimization task across different RL paradigms. Among conventional RL baselines, advanced actor-critic methods (e.g., TD3 and SAC) achieve competitive performance by improving critic stabilization and enhancing exploration. By contrast, Euclidean Gaussian does not consistently surpass these strong RL baselines. Although diffusion-style sampling introduces multi-step stochastic exploration, Euclidean Gaussian perturbations treat RIS phases as unconstrained real vectors and thus fail to account for the periodicity of $[0,2\pi)$, which can yield geometrically distorted intermediate candidates and reduce sampling efficiency in high-dimensional phase spaces. 

Compared with the proposed RDM, the RDM achieves the best asymptotic reward and faster convergence. This superiority directly stems from its phase-tailored noise design and manifold-consistent denoising. By injecting and removing noise on the torus manifold $(\mathbb{S}^{1})^{N}$ and preserving phase periodicity throughout the reverse process, RDM maintains feasibility at every denoising step and generates higher-quality phase candidates for critic-guided refinement. These results confirm the advantage of RDM for high-dimensional RIS phase.

\subsection{Impact of Model Configurations}
\begin{figure}[t]
    \centering
    \captionsetup{font={small}}
    \includegraphics[width=1\columnwidth]{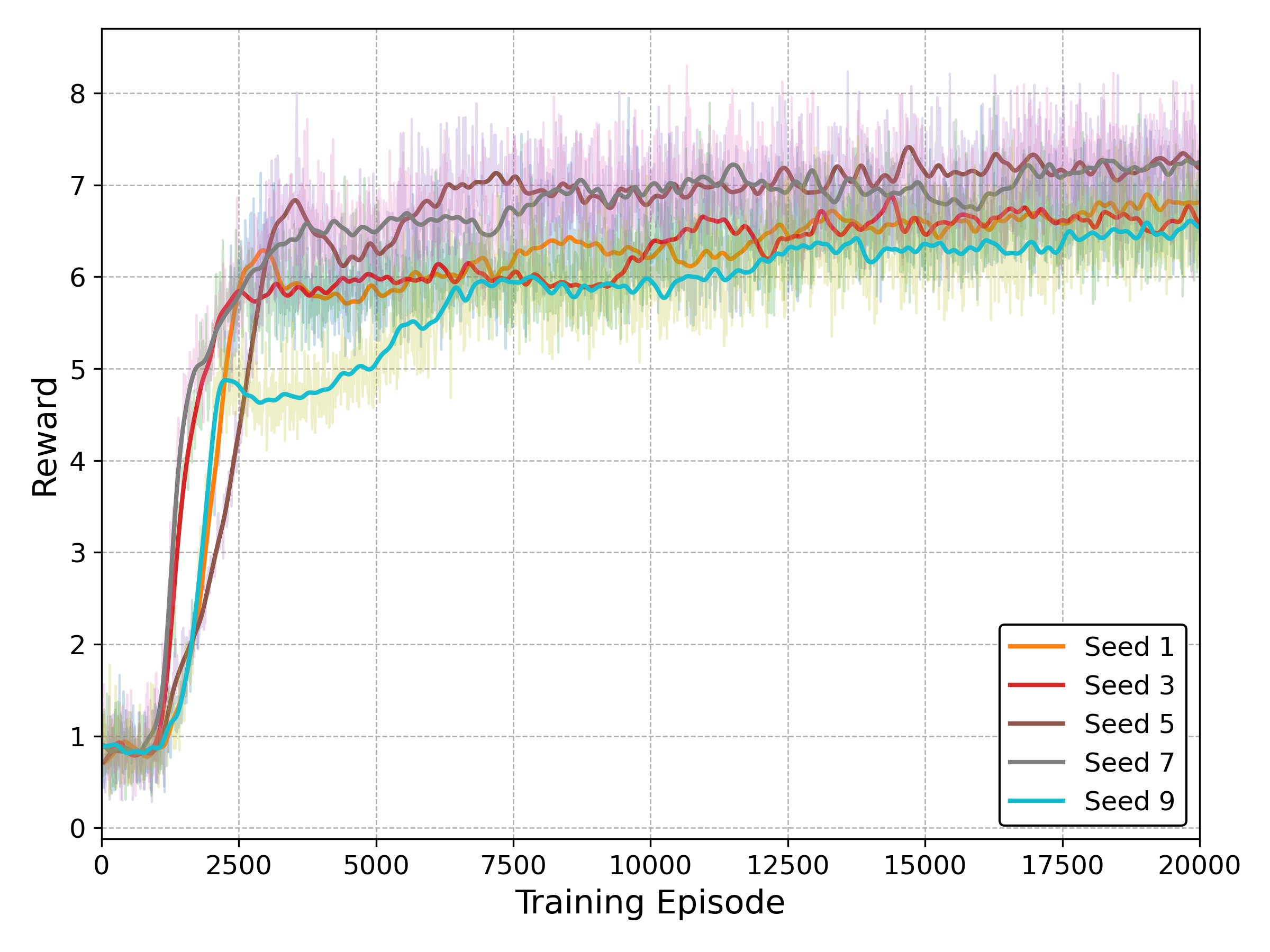}
    \caption{Learning curves of the proposed RDM-DRL framework under different random seeds, illustrating training stability and convergence behavior.}
    \label{fig:seed}
\end{figure}

To assess the sensitivity and stability of the proposed diffusion-DRL framework, we conduct a series of parameter studies focusing on three key components of the generative policy, e.g., the random seed, the noise schedule, and the number of denoising steps. These factors influence different stages of the learning pipeline, from stochastic initialization and exploration to the forward noise process structure to the reverse denoising refinement depth. Understanding their effects is crucial for ensuring stable training and robust RIS phase optimization.

\subsubsection{Impact of Random Seeds}
Initially, we examine the sensitivity of the proposed diffusion-DRL framework to random seed choices. Since reinforcement learning involves stochastic elements in network initialization, environment sampling, and exploration noise, different seeds can yield significant variation in learning behavior. To assess this effect, we train the RDM-based agent under five distinct seeds while keeping all other configurations fixed. As depicted in Fig.~\ref{fig:seed}, all training curves exhibit stable growth and eventually converge to high-reward regions, although the transient learning trajectories vary across seeds. This consistency indicates that the proposed RDM architecture remains robust to initialization and stochasticity and is less susceptible to collapse or divergence, a common issue in conventional DRL pipelines. The results confirm that the manifold-aware diffusion process provides a more stable exploration mechanism, effectively reducing seed-induced variability.

\subsubsection{Impact of Noise Schedule Functions}
\begin{figure}[t]
    \centering
    \captionsetup{font={small}}
    \includegraphics[width=1\columnwidth]{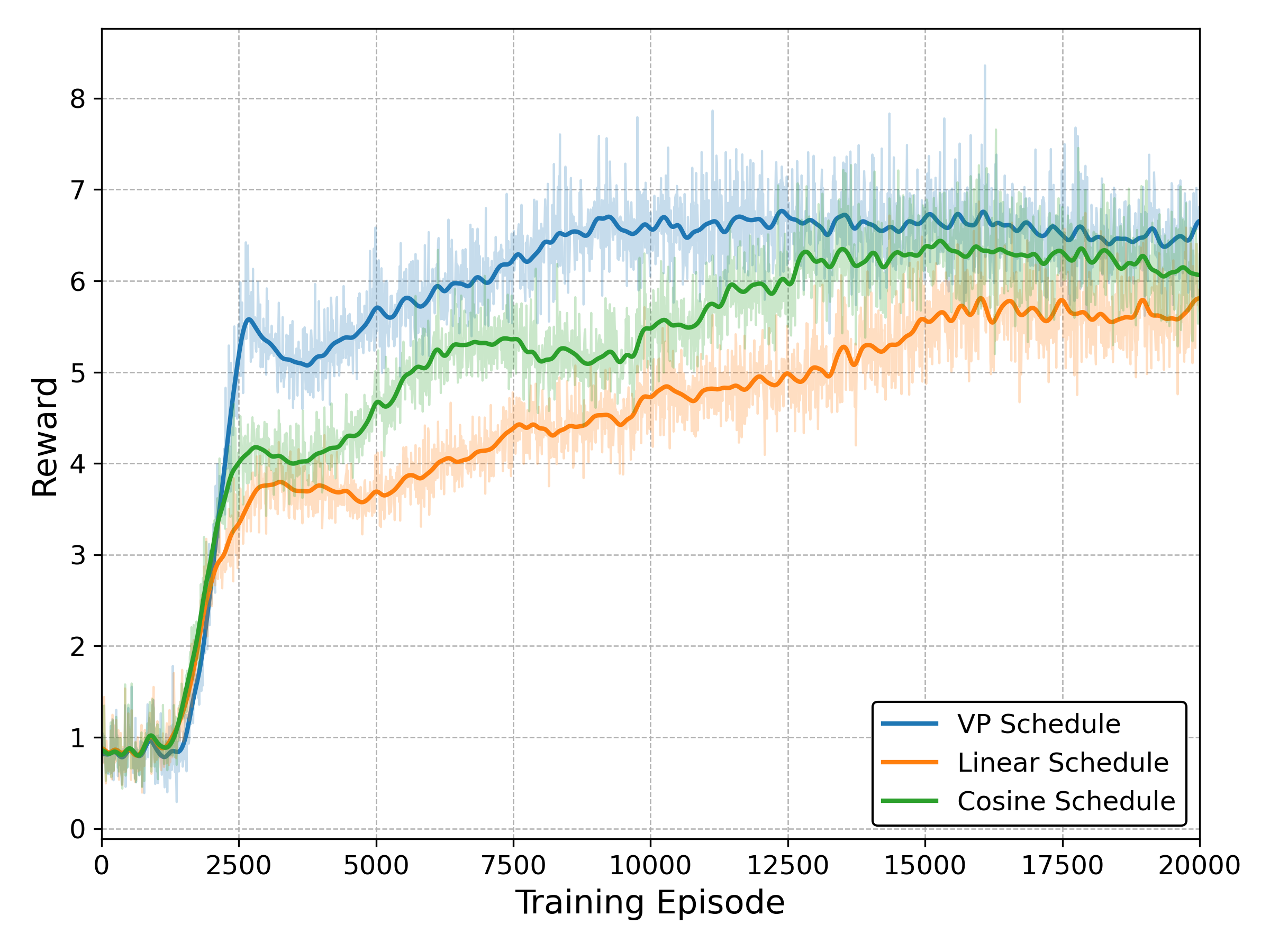}
    \caption{Performance comparison of different diffusion noise schedules (VP, linear, and cosine) in terms of learning dynamics and final reward.}
    \label{fig:schedule}
\end{figure}

Next, we investigate how different noise scheduling strategies affect the performance of the diffusion model. The noise schedule governs the denoising strength and thus affects the complexity of the reverse denoising task. We compare three widely used schedules, variance-preserving (VP), linear, and cosine, within identical training configurations. Fig.~\ref{fig:schedule} shows the resulting learning curves. The VP schedule consistently achieves the highest average reward and converges faster, while the linear schedule shows moderate performance and the cosine schedule lags. This behavior suggests that VP noise yields a more uniform distribution of noise levels across the diffusion horizon, enabling the network to learn a more accurate approximation of the phase-manifold distribution. In contrast, overly shallow or overly aggressive schedules introduce mismatched gradients during reverse sampling, which reduces training efficiency.

\subsubsection{Impact of Denoising Steps}
\begin{figure}[t]
    \centering
    \captionsetup{font={small}}
    \includegraphics[width=1\columnwidth]{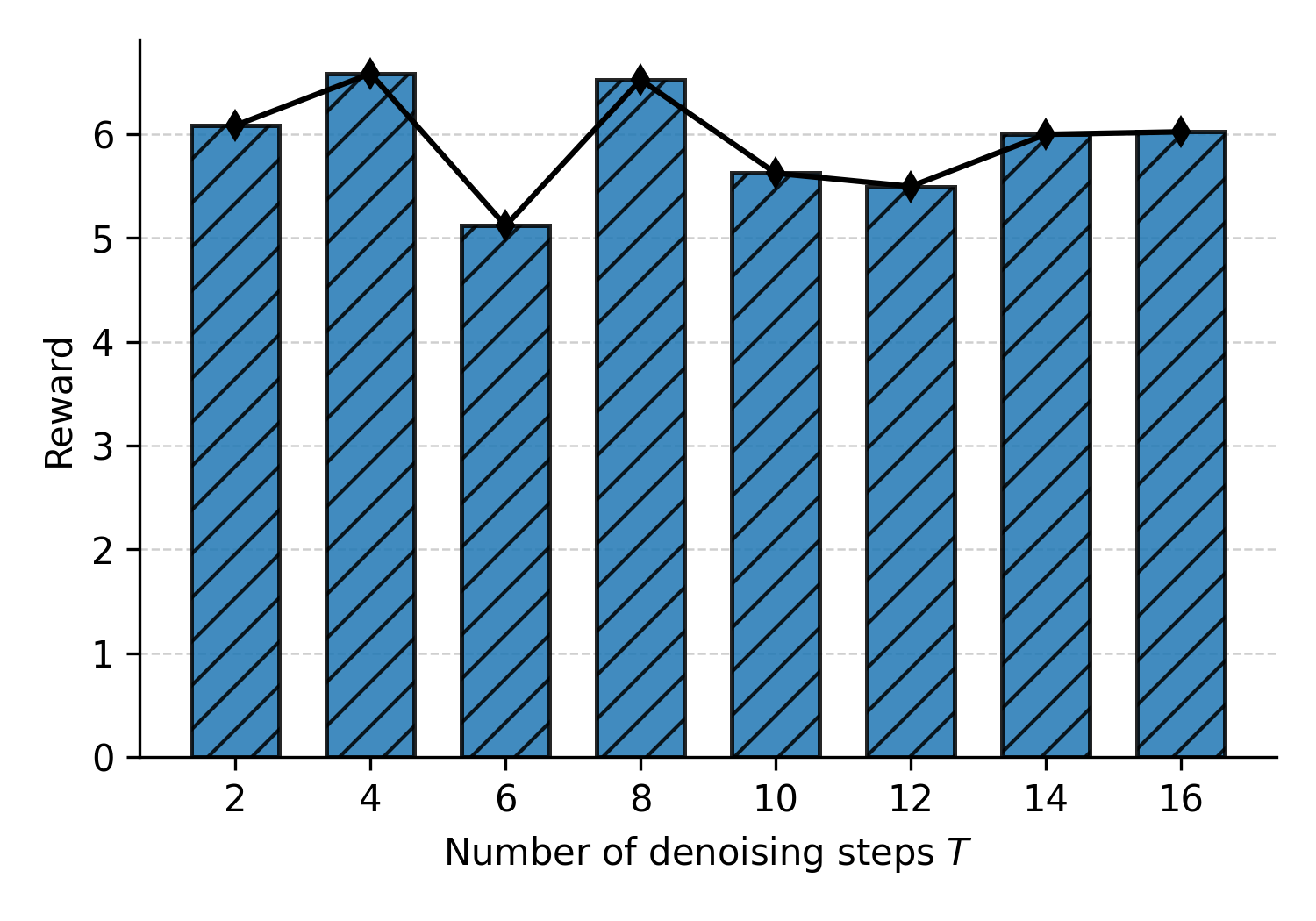}
    \caption{Impact of the number of denoising steps on the final achievable reward of the proposed diffusion-DRL policy.}
    \label{fig:denoisingStep}
\end{figure}

Finally, we examine the effect of the number of denoising iterations $T$ in the reverse diffusion process. The denoising depth dictates how thoroughly the model refines noisy samples and is closely linked to computational cost. We evaluate $T \in \{2, 4, 6, 8, 10, 12, 14, 16 \}$ using the same VP schedule. As shown in Fig.~\ref{fig:denoisingStep}, performance initially improves as $T$ increases, with $T=4$ yielding the highest final reward. However, further increasing the number of denoising steps results in diminishing returns and, in some cases, noticeable degradation. This pattern arises because excessively long reverse chains may overfit to intermediate noise patterns or introduce redundant corrections that accumulate and distort the manifold-consistent phase updates. The results emphasize the importance of selecting an appropriate denoising depth that balances refinement accuracy and stability, with moderate step sizes providing the best trade-off for RIS phase optimization.

\subsection{Impact of Communication Parameters}

\begin{figure}
    \centering
    \captionsetup{font={small}}
    \includegraphics[width=1\columnwidth]{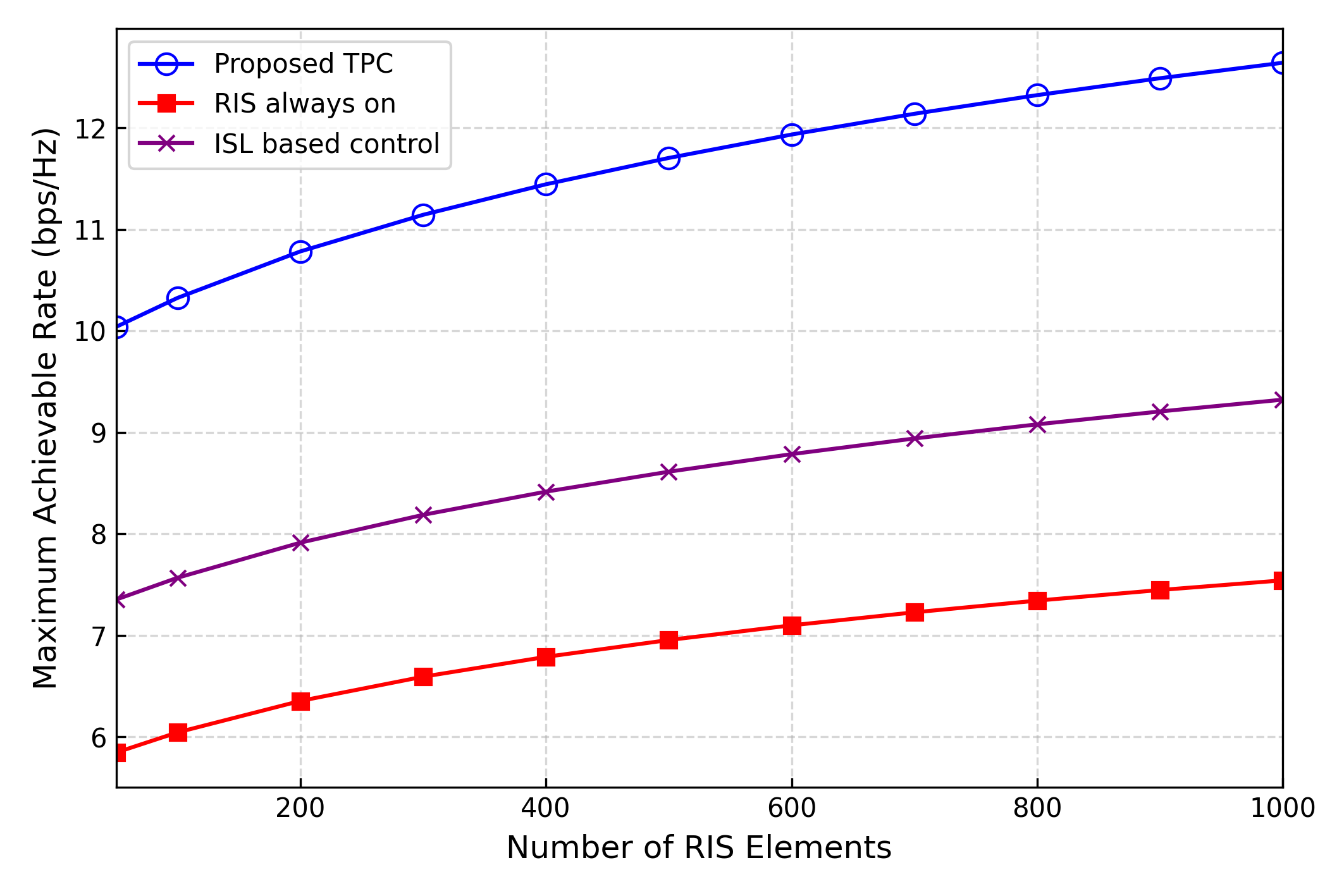}
    \caption{SINR performance of the proposed framework under different numbers of RIS reflecting elements.}
    \label{fig:element_analysis}
\end{figure}

\begin{figure}[t]
    \centering
    \includegraphics[width=0.9\linewidth]{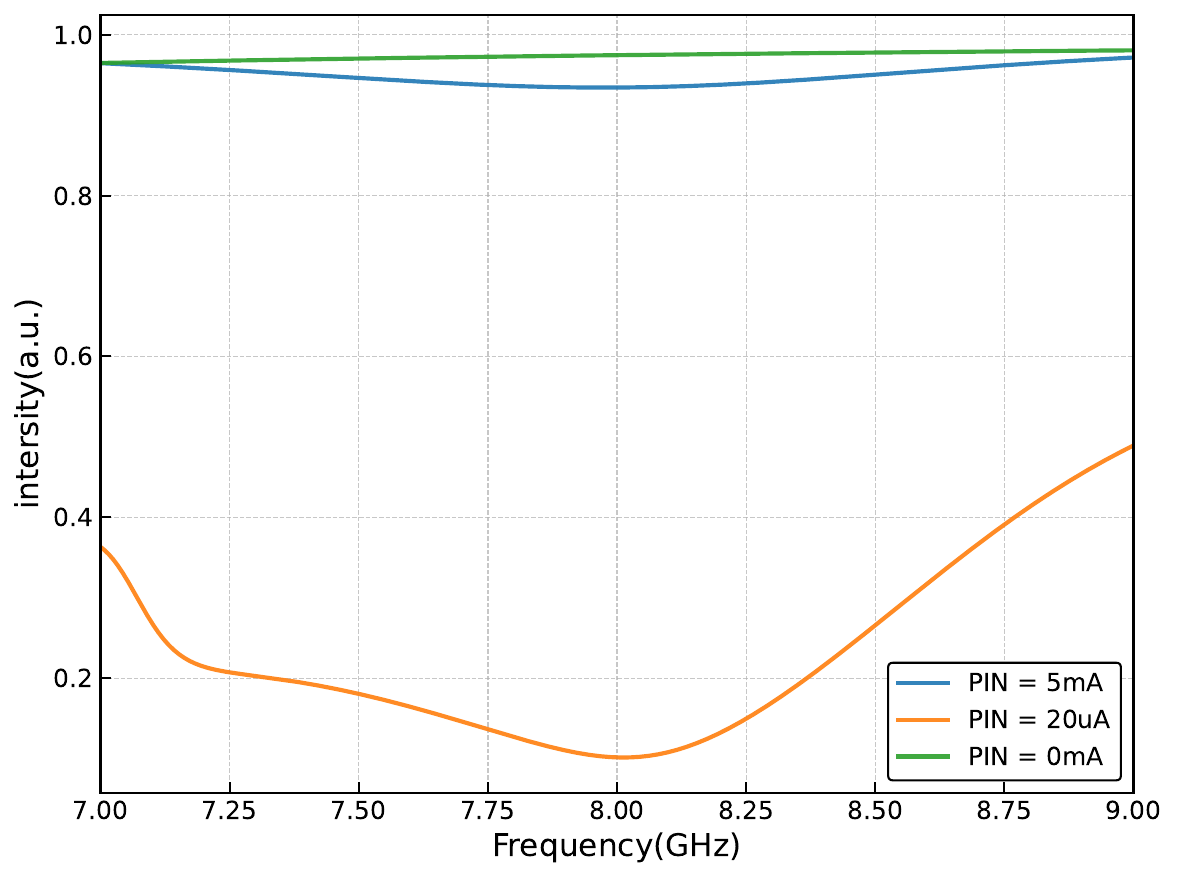}
    \caption{Reflection intensity of the proposed RIS unit cell under different PIN-diode bias currents.}
    \label{fig:magnitude_comparison}
\end{figure}

Fig.~\ref{fig:element_analysis} shows the variation in SINR across different configurations of RIS element numbers. Overall, as the number of RIS units increases, signal enhancement improves, and SINR increases. However, we also observe that when the number of RIS units is large, irrational reflections can exacerbate interference, leading to significant spectral pollution. The proposed TPGC algorithm effectively mitigates interference enhancement from invalid reflections via a reasonable ON/OFF control strategy, ultimately achieving optimal SINR performance.

\begin{figure}[t]
    \centering
    \includegraphics[width=0.9\linewidth]{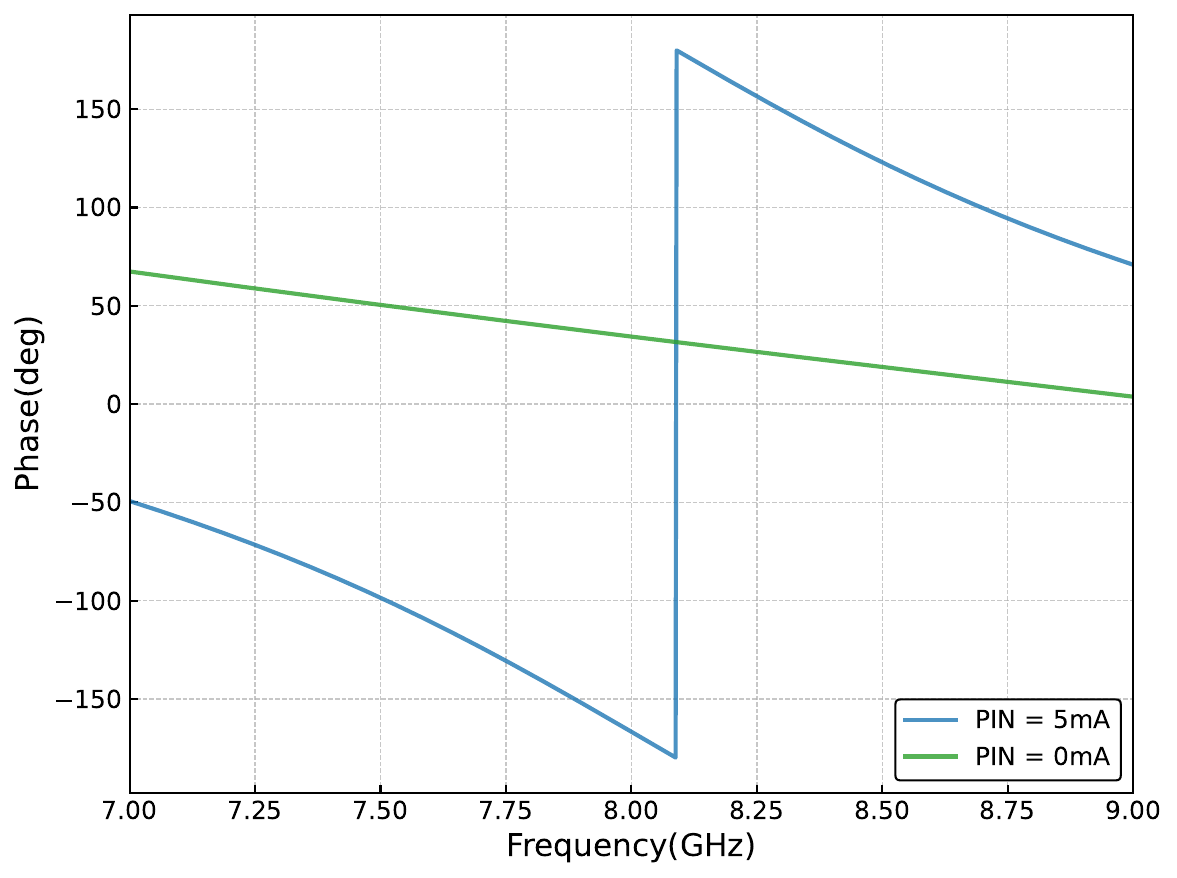}
    \caption{Phase response of the proposed RIS unit cell under different PIN-diode bias currents.}
    \label{fig:phase_comparison}
\end{figure}

\subsection{Implementation of ON/OFF states}
To validate the hardware-consistent ON/OFF switching behavior and phase programmability, we perform full-wave electromagnetic simulations of the proposed unit cell in CST Microwave Studio.
We focus on the operating frequency of $8$~GHz and evaluate the reflected-signal strength, where the vertical axis in Fig.~\ref{fig:magnitude_comparison} is reported as the reflection intensity (a.u.).
As shown in Fig.~\ref{fig:magnitude_comparison}, when the RIS is in the ON state with the diode bias current $PIN\in\{0~\mathrm{mA},\,5~\mathrm{mA}\}$, the reflection intensity is close to $1$, indicating that the incident wave is almost fully reflected.
In contrast, when the RIS is switched to the OFF state with $PIN=20~\mu\mathrm{A}$, the reflection intensity approaches zero at $8$~GHz, which confirms that the incident signal is largely absorbed and the reflected component is effectively suppressed.

Moreover, Fig.~\ref{fig:phase_comparison} demonstrates the phase programmability in the ON state. By adjusting the diode bias current (e.g., from $0$~mA to $5$~mA), the reflected phase at $8$~GHz can be tuned to realize an approximately $180^\circ$ phase shift.
These CST results support our modeling assumption that $v_i=1$ enables reflective beamforming with controllable phase, whereas $v_i=0$ realizes an absorptive mode that effectively removes the RIS interference.

\section{Conclusion}\label{s8}
This paper presents a trajectory-aware generative ON/OFF control framework. We designed this system specifically for multi-RIS-assisted uplink communications in highly dense mobile environments. Through leveraging an enhanced LSTM-based trajectory predictor, the BS can anticipate future user mobility and reconstruct the corresponding CSI of RIS-parametrized channels, enabling proactive, interference-aware RIS configuration. We developed an RDM operating on the torus manifold to address the mixed discrete-continuous and non-convex nature of RIS control. We integrated with DRL-based guidance to generate geometry-consistent RIS phase configurations and effective ON/OFF control decisions. Simulation results using real-world user mobility data showcased our framework's performance. The proposed framework consistently outperforms representative RIS control strategies and learning-based baselines. It excels in SINR performance, mobility robustness, and generalization across different interference conditions. These findings confirm the value of the proposed TPGC approach. It provides a promising framework for reliable multi-RIS-assisted communications in large-scale mobile events.

\bibliographystyle{IEEEtran}
\bibliography{IEEEabrv, reference}

\end{document}